\documentclass[11pt, a4paper, logo, copyright, nonumbering]{seed}

\usepackage{natbib}

\usepackage{CJKutf8}

\usepackage{xargs}  

\usepackage{todonotes}  

\usepackage{multirow}

\usepackage{cleveref}

\usepackage{amsmath}
\usepackage{dsfont}

\usepackage{subcaption}

\usepackage{svg}

\usepackage{listings}
\lstdefinestyle{mystyle}{
    commentstyle=\color{green!50!black},
    numberstyle=\tiny\color{gray},
    basicstyle=\ttfamily\fontsize{5}{5}\selectfont,
    breakatwhitespace=false,         
    breaklines=true,                 
    captionpos=b,                    
    keepspaces=true,                 
    numbers=none,                    
    numbersep=5pt,                  
    showspaces=false,                
    showstringspaces=false,
    showtabs=false,                  
    tabsize=2
}

\lstdefinestyle{mystyle2}{
    commentstyle=\color{green!50!black},
    numberstyle=\tiny\color{gray},
    basicstyle=\ttfamily\fontsize{7}{7}\selectfont,
    breakatwhitespace=false,         
    breaklines=true,                 
    captionpos=b,                    
    keepspaces=true,                 
    numbers=none,                    
    numbersep=5pt,                  
    showspaces=false,                
    showstringspaces=false,
    showtabs=false,                  
    tabsize=2
}
\usepackage{makecell}
\usepackage{xspace}
\usepackage{tabularx}
\usepackage{array}
\usepackage{wrapfig}
\usepackage{subcaption}
\usepackage{multirow}
\usepackage{rotating} 
\usepackage{multicol}
\usepackage{setspace}
\usepackage{adjustbox} 
\usepackage{tipa}

\usepackage{graphicx}
\usepackage{hyperref}
\usepackage[most]{tcolorbox}
\definecolor{mygray}{gray}{0.9}
\definecolor{mydarkblue}{rgb}{0,0.08,0.45}
\hypersetup{
 colorlinks=true,
 citecolor=mydarkblue,
 urlcolor=mydarkblue
}
\usepackage{footmisc}

\newcommand{\creditsectionheader}[1]{\parbox{\columnwidth}{\centering \textbf{\large #1}}\\}

\newcommand{\corecontributor}[1]{#1\\}
\newcommand{\multiswebench}{Multi-SWE-bench\xspace}
\newcommand{\multiswerl}{Multi-SWE-RL\xspace}
\newcommand{\agentless}{Agentless\xspace}
\newcommand{\magentless}{MagentLess\xspace}
\newcommand{\sweagent}{SWE-agent\xspace}
\newcommand{\msweagent}{MSWE-agent\xspace}
\newcommand{\openhands}{OpenHands\xspace}
\newcommand{\mopenhands}{MopenHands\xspace}

\renewcommand{\today}{}
\newcommandx{\info}[2][1=]{\todo[linecolor=red,backgroundcolor=red!25,bordercolor=red,#1]{#2}}

\title{\centering \multiswebench: A Multilingual Benchmark \\ for Issue Resolving}

\author[*]{
\vspace{-10pt}
\textbf{ByteDance Seed}
\vspace{0pt}
\\
\noindent
\makebox[\textwidth]{
  \raisebox{-0.2ex}{%
    \tcbox[colback=white, colframe=white, left=1pt, right=1pt, boxrule=0pt, arc=0mm]{
      \includegraphics[height=1em]{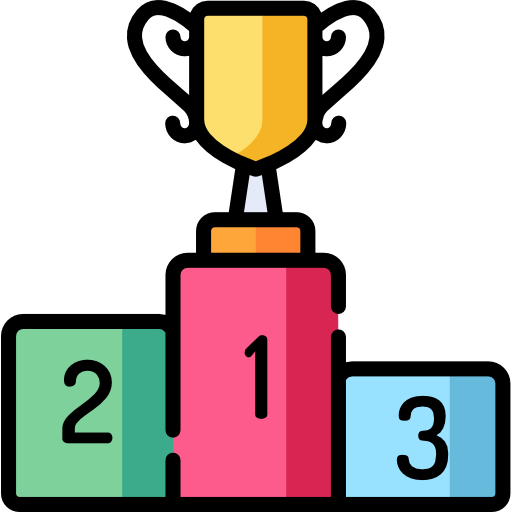}\,
      \href{https://multi-swe-bench.github.io}{Leaderboard}
    }
  }
  \hspace{-0.5em}
  \raisebox{-0.2ex}{%
    \tcbox[colback=white, colframe=white, left=1pt, right=1pt, boxrule=0pt, arc=0mm]{
      \includegraphics[height=1em]{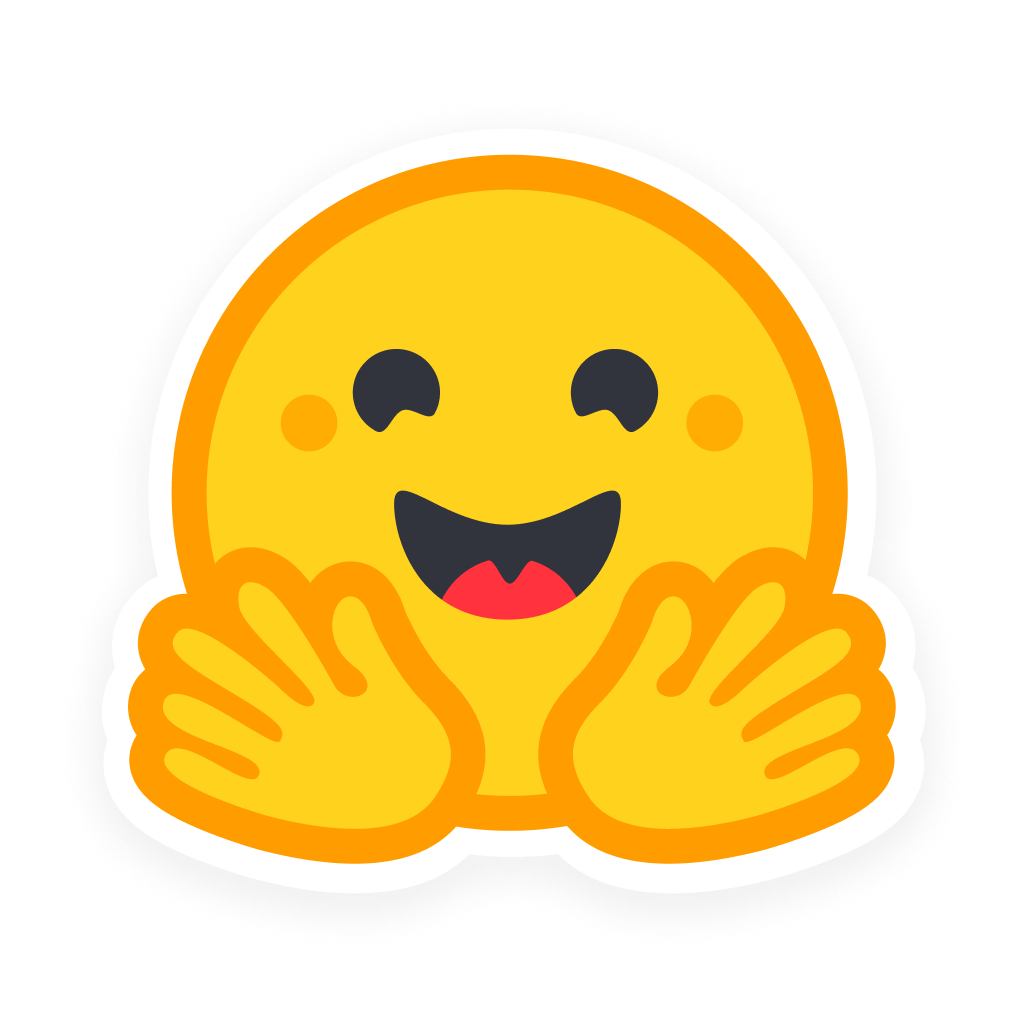}\,
      \href{https://huggingface.co/datasets/bytedance-research/Multi-SWE-bench}{Benchmark}
    }
  }
  \hspace{-0.5em}
  \raisebox{-0.8ex}{%
    \tcbox[colback=white, colframe=white, left=1pt, right=1pt, boxrule=0pt, arc=0mm]{
      \includegraphics[height=1em]{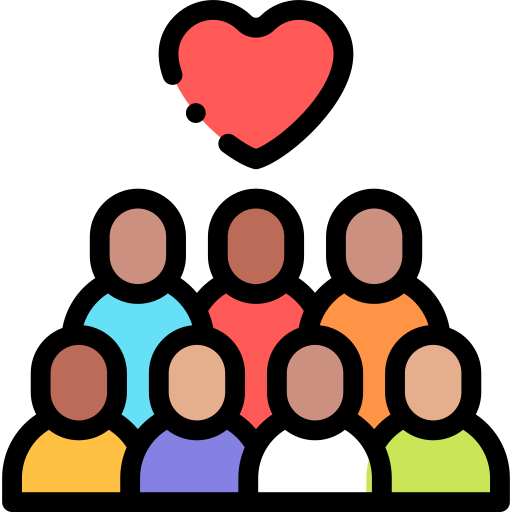}\,
      \href{https://huggingface.co/datasets/bytedance-research/Multi-SWE-RL}{RL Community}
    }
  }
  \hspace{-0.5em}
  \raisebox{-0.9ex}{%
    \tcbox[colback=white, colframe=white, left=1pt, right=1pt, boxrule=0pt, arc=0mm]{
      \includegraphics[height=1em]{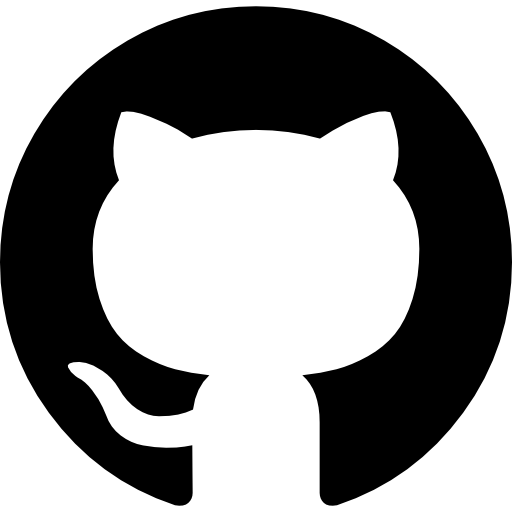}\,
      \href{https://github.com/multi-swe-bench/multi-swe-bench}{GitHub Repo}
    }
  }
}
\vspace{-30pt}
}

\begin{document}

\begin{abstract}
\vspace{-15pt}
The task of issue resolving is to modify a codebase to generate a patch that addresses a given issue.
However, existing benchmarks, such as SWE-bench, focus almost exclusively on Python, making them insufficient for evaluating Large Language Models (LLMs) across diverse software ecosystems.
To address this, we introduce a multilingual issue-resolving benchmark, called \multiswebench, covering Java, TypeScript, JavaScript, Go, Rust, C, and C++.
It includes a total of $1,632$ high-quality instances, which were carefully annotated from $2,456$ candidates by $68$ expert annotators, ensuring that the benchmark can provide an accurate and reliable evaluation.
Based on \multiswebench, we evaluate a series of state-of-the-art models using three representative methods (\agentless, \sweagent, and \openhands) and present a comprehensive analysis with key empirical insights.
In addition, we launch a \multiswerl open-source community, aimed at building large-scale reinforcement learning (RL) training datasets for issue-resolving tasks.
As an initial contribution, we release a set of $4,723$ well-structured instances spanning seven programming languages, laying a solid foundation for RL research in this domain.
More importantly, we open-source our entire data production pipeline, along with detailed tutorials, encouraging the open-source community to continuously contribute and expand the dataset.
We envision our \multiswebench and the ever-growing \multiswerl community as catalysts for advancing RL toward its full potential, bringing us one step closer to the dawn of AGI.
\end{abstract}

\maketitle

\let\oldthefootnote\thefootnote
\renewcommand*{\thefootnote}{\fnsymbol{footnote}}
\footnotetext[0]{Author contributions \hyperref[sec:contributions]{listed at end of paper}. Correspondence to: \{zandaoguang, shen.kai\}@bytedance.com}
\let\thefootnote\oldthefootnote

\vspace{-0.5em}
\begin{figure}[h]
    \centering
    \includegraphics[width=0.90\linewidth]{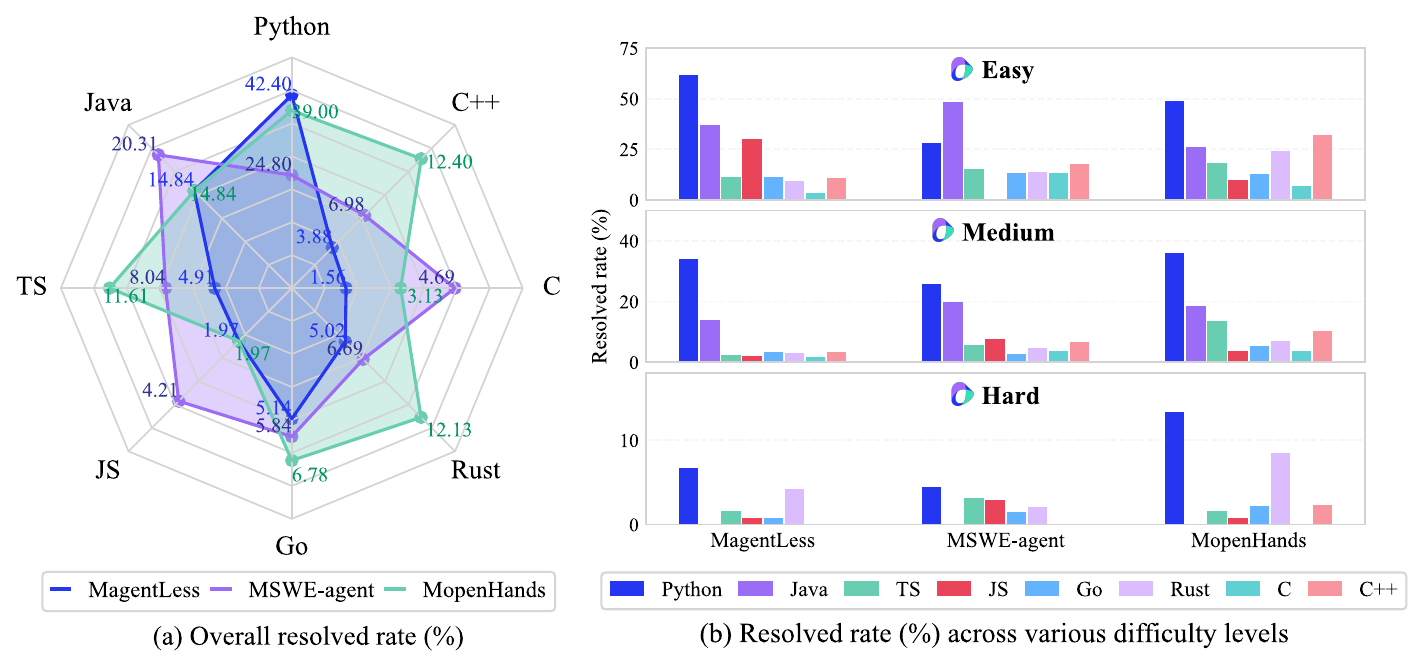}
    \caption{Resolved rate (\%) on \multiswebench (Claude-3.5-Sonnet).}
    \label{fig:radar_diagram}
\end{figure}

\newpage

\tableofcontents

\newpage

\section{Introduction}

Automating software engineering tasks with large language models (LLMs) has gained considerable attention~\citep{nl2code,zheng2023survey,survey_2,survey_3} recently.
Beyond code generation, the issue resolving task proposed by SWE-bench~\citep{swe-bench} changes the role of LLMs from code assistants to fully autonomous AI programmers.
SWE-bench contains $2,294$ issues from $12$ widely-used open-sourced Python libraries.
LLMs are tasked to generate a patch based on the issue description along with the buggy code repository.
SWE-bench Verified is a subset of $500$ human-validated issues selected from SWE-bench, chosen for appropriately scoped unit tests and well-specified issue descriptions.
Within less than one year, the resolving rate on SWE-bench Verified increased from $0.40\%$~\citep{swe-bench} (for RAG+GPT3.5) to $65.40\%$~\citep{augmentagentv0} (for Augment Agent v0).

Although existing works based on SWE-bench demonstrate significant progress in Python-based issue resolving, the diversity of programming languages in real-world repositories presents additional challenges that remain unexplored. 
In particular, repositories in different languages follow distinct programming paradigms, idiomatic patterns, and runtime behaviors, which may impact the effectiveness of current approaches. 
This raises the question of whether the impressive performance of existing agents on Python issues can be generalized to other widely used languages, such as Java, TypeScript, JavaScript, Go, Rust, C, and C++.

To answer this question, we introduce \multiswebench, a multilingual benchmark for issue resolving, consisting of $1,632$ issues across $7$ widely used programming languages: Java, TypeScript, JavaScript, Go, Rust, C, and C++.
To construct a reliable benchmark for evaluating the ability of agents to resolve real-world software issues, we employ a systematic five-phase pipeline. 
First, we select high-quality repositories from GitHub based on star ratings and runnability counts to ensure both popularity and practical usability. 
Second, we collect issue-related pull requests (PRs) along with their corresponding metadata. 
Third, we build Dockerized environments for each PR by extracting dependencies from CI/CD workflows and documentation to ensure reproducible execution.
Fourth, we validate PRs by analyzing test outcomes across patch configurations, retaining only those with clear bug-fixing effects and no regressions.
Fifth, we perform rigorous manual verification through dual annotation and cross-review, ensuring high-quality ground truth aligned with SWE-bench verified standards.
By ensuring diversity, executability, and human-verified correctness, \multiswebench sets a high standard for evaluating LLMs on realistic and non-trivial issue-resolving tasks.


With its wide coverage of languages and issue types, \multiswebench introduces realistic challenges that push the boundaries of LLM-based software agents.
We use \multiswebench to evaluate the generalizability of 3 representative methods (i.e., \agentless~\citep{xia2024agentless}, \sweagent~\citep{yang2024sweagent}, and \openhands+CodeAct v2.1~\citep{openhands}) based on 9 top-performing frontier models (i.e., GPT-4o, OpenAI-o1, OpenAI-o3-mini-high, Claude-3.5-Sonnet, Claude-3.7-Sonnet, DeepSeek-V3, DeepSeek-R1, Qwen2.5-72B-Instruct, and Doubao-1.5-Pro).
Our evaluation provides a comparative analysis of the overall effectiveness of these methods across seven programming languages, along with Python, offering insights into their cross-language capabilities. 
Furthermore, we conduct a fine-grained analysis of the key factors influencing model performance and investigate failure cases for each language to identify underlying challenges and limitations.
Through comprehensive analysis and comparison, we provide a good understanding of existing models and shed light on future directions and further progress.
For example, our findings show that models perform generally better when issue descriptions are longer, indicating a strong reliance on rich contextual grounding; 
in contrast, resolved rates drop sharply when fix patches exceed $600$ tokens or touch more than one file, exposing weaknesses in long-context retention and multi-file reasoning.
Together, these findings delineate the current boundary of LLM capabilities in software engineering and define the key obstacles to real-world deployment.

Beyond \multiswebench, we launch the \multiswerl open-source community to address the pressing need for scalable, high-quality RL environments in software engineering.
Recent models such as DeepSeek-R1~\citep{deepseekr1}, OpenAI-o1~\citep{openaio1}, and OpenAI-o3~\citep{openai_o3_mini} have demonstrated the potential of RL even with simplistic reward signals.
These advances reinforce our belief that "\textit{scaling RL in real-world software environments is a key pathway toward human-level intelligence}".
However, the creation of realistic, interactive environments remains a major bottleneck.
As a first step toward scalable RL in software engineering, \multiswerl therefore launches a collaborative initiative to build training data and environment from real-world tasks.
As the initial contribution to the \multiswerl community, we release a dataset of $4,723$ containerized issue-resolving instances spanning $7$ programming languages.
Each instance is equipped with a reproducible execution environment, enabling plug-and-play training for RL agents in realistic software contexts.
We envision this release as a spark—igniting broader community interest in the construction of RL training data 
and paving the way toward fully autonomous agent systems.

In summary, our main contributions are:
\begin{itemize}
    \item \multiswebench, a multilingual benchmark for issue resolving, consisting of $1,632$ human-validated GitHub issues on $7$ widely used programming language. It serves as a reliable support for comprehensively evaluating the performance of agents in real-world software development scenarios.
    \item A large-scale evaluation of $9$ state-of-the-art LLMs across $3$ representative methods on \multiswebench, yielding diagnostic insights to guide future research.
    \item \multiswerl, a community-driven open-source effort that initiates scalable RL data creation from real-world software tasks, laying the groundwork for long-term progress in multilingual software agent development.
\end{itemize}

\section{Related Work}
The remarkable performance of LLMs in code-related tasks has motivated substantial research to study their role in automating software engineering.
To evaluate the capabilities and limitations of existing approaches, a wide range of benchmarks for code-related tasks has been developed.
Early efforts in this domain focused on primarily evaluating models in monolingual program-level evaluations~\citep{allamanis2013mining, raychev2016probabilistic, iyer2018mapping,chen2021evaluating, austin2021program, wang2023execution}.
As LLMs advanced, benchmarks evolved in two key dimensions to better align with real-world software engineering scenarios. 
First, benchmarks shift from monolingual to multilingual tasks, with growing interest and practical needs in evaluating LLMs' performance across multiple programming languages. 
Examples include Multilingual-HumanEval~\citep{athiwaratkun2023multi} and HumanEval-X~\citep{zheng2023codegeex}, which extend the HumanEval~\citep{codex} benchmark to multiple languages, and MBXP~\citep{athiwaratkunmulti}, which extends MBPP to multilingual scenarios. 
Second, benchmarks shift from program-level to repository-level tasks, focusing on more complex scenarios such as library-oriented code generation~\citep{zan2022cert}, repository-level code completion~\citep{zhang2023repocoder,liurepobench,ding2024crosscodeeval,graphcoder,yu2024codereval}, and bug fix~\citep{mundler2024swt, ouyang2024benchmarking, sunrepofixeval,saavedra2024gitbug}. 
These evolving benchmarks aim to provide a more comprehensive evaluation of LLMs in real-world software development environments.

In addition to existing benchmarks, SWE-bench~\citep{swe-bench} has gained significant attention since its release.
Instead of focusing on isolating code subtasks into separate datasets, SWE-bench addresses a broader range of tasks through repository-level issue resolving.
These issue resolving tasks, including bug fixing, new feature requests, and optimization, which provide a more comprehensive evaluation of LLMs' ability to automating software development.
While SWE-bench is limited to textual context, SWE-bench Multimodal~\citep{yang2024swebenchmultimodal} and Visual SWE-bench~\citep{codev} extend evaluation to systems fixing bugs in visually-oriented and user-facing applications.
SWE-Lancer~\citep{swelancer} focuses on JavaScript and TypeScript, featuring over $1,400$ freelance tasks from Upwork, including technical and managerial tasks.
Despite these advancements, the performance of LLMs on other widely used programming languages remains underexplored. 
Our work aims to bridge this gap with \multiswebench, a large-scale multilingual benchmark for issue resolving, with $1,632$ human-validated GitHub issues across $7$ widely used languages.

\section{\multiswebench}
\label{sec:multiswebench}

\multiswebench consists of $1,632$ issue-resolving tasks spanning $7$ programming languages: Java, TypeScript, JavaScript, Go, Rust, C, and C++.
This section will provide the construction process of \multiswebench, along with an analysis of its key features. 

\subsection{Benchmark Construction}
\label{sec:workflow_overview}

\begin{figure*}
    \centering
    \includegraphics[width=0.98\linewidth]{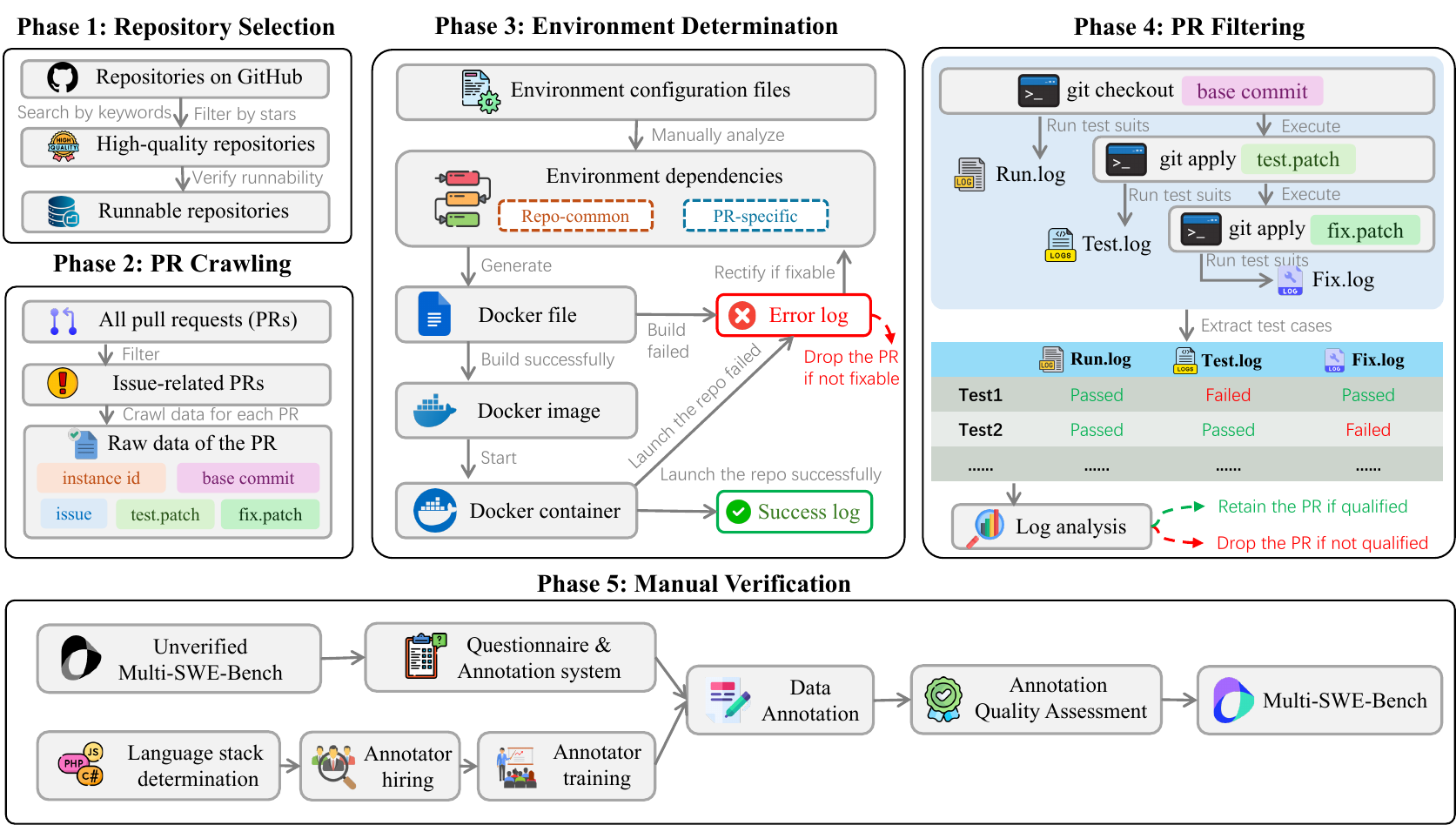}
    \vspace{-3mm}
    \caption{Construction of \multiswebench.}
    \label{fig:dataset_construction}
\end{figure*}

To evaluate the generalizability of LLMs as issue resolvers, seven programming languages are selected to construct \multiswebench through five phases.
As shown in Fig.~\ref{fig:dataset_construction}, the first four phases create a large pool of candidate data for each language, while the fifth phase finalizes the \multiswebench through manual verification.

\subsubsection{Phase 1: Repository Selection}
\label{sec:phase1_repository_selection}

We carefully curate a diverse set of high-quality GitHub repositories for each of the seven target programming languages. 
The selection process is guided by the following criteria:
\begin{itemize}[leftmargin=*]
    \item Popularity and Maintenance: Repositories must have over $500$ GitHub stars and demonstrate active maintenance for at least six months. 
    In addition, we prioritize repositories frequently recommended in Google searches using keywords such as "high-quality", "well-maintained", and "popular".
    \item CI/CD Support: Selected repositories are required to include CI/CD configurations (e.g., workflows under "\texttt{.github/workflows/}") to ensure automated testing and reproducibility.
    \item Build Viability: After minimal manual setup, the latest commit must be buildable and testable in a clean environment, ensuring compatibility with modern tooling and infrastructure.
\end{itemize}
This filtering process results in a robust foundation of repositories that are representative of real-world, production-level codebases, laying the groundwork for subsequent phases.

\subsubsection{Phase 2: Pull Request Crawling}
\label{sec:phase2_pull_request_crawling}

This phase aims to crawl issue-resolving pull requests (PRs) for each repository selected in phase 1.
All PRs from the repository are collected and then filtered based on the following criteria:
\begin{itemize}[leftmargin=*]
    \item Linked with at least one GitHub issue: The PR must be linked to at least one issue to ensure it addresses a clearly defined bug report or feature request.
    \item Modified test files: The PR must include changes to test files, guaranteeing proper testing is in place to verify the correctness of the fix patches.
    \item Merged into the main branch: The PR must be merged into the main branch, indicating it has been accepted by the repository's maintainers and fully integrated.
\end{itemize}
After filtering, detailed information is gathered for each PR, including attributes such as issue description, base commit, fix.patch, and test.patch.

\subsubsection{Phase 3: Environment Determination}



To enable faithful execution and evaluation of issue-resolving tasks, each PR must be reproducibly built and executed in an isolated environment.
In this phase, we construct a Docker-based runtime environment for every PR by automatically identifying and provisioning its necessary dependencies.
The process begins with a manual inspection of environment-related artifacts, including CI/CD configuration files (e.g., GitHub Actions), repository documentation (e.g., README files), and exploratory trial runs.
Through this analysis, we extract environment dependencies and classify them into two categories: repo-common dependencies, which are shared across the entire repository, and PR-specific dependencies, which are introduced or modified by the target PR.

Using the extracted dependency information, we generate a tailored Dockerfile and attempt to build a corresponding Docker image.
If the build fails, we examine the error logs to identify missing dependencies, misconfigurations, or version conflicts. For fixable errors, we iteratively patch the Dockerfile or supporting scripts; otherwise, we discard the PR to ensure reliability.
Once the image is successfully built, we verify whether the repository can be launched at the specific commit associated with the PR. 
This step ensures that all required services, packages, and configurations are functional.
If the launch fails, we again attempt corrective actions; if successful, we obtain a validated and executable containerized environment for downstream evaluation.
This phase ensures that each PR is equipped with a clean and functional containerized environment, laying a necessary foundation for subsequent testing and analysis.

\subsubsection{Phase 4: Pull Request Filtering}

With both the PR metadata and a working runtime environment established in the previous phases, we now perform semantic validation to ensure each PR meets the requirements of issue resolving. 
This is done by analyzing test behaviors under controlled patch configurations.
For each PR, unlike SWE-bench which runs only relevant tests, we run the full test suit under the following three settings:
\begin{itemize}[leftmargin=*]
    \item Run.log: Tests are executed on the base commit.
    \item Test.log: The test.patch is applied to the base commit before execution.
    \item Fix.log: Both the test.patch and the fix.patch are applied to the base commit before execution.
\end{itemize}
We then extract the execution status of each test case from these logs.
Unlike SWE-bench, which considers only \texttt{PASSED} and \texttt{FAILED}, we also track \texttt{NONE} and \texttt{SKIPPED} status, as some test cases may be conditionally disabled or omitted after applying patches—resulting in inconsistent test counts across the three logs.
Each test case is summarized by its status transition across the three settings. 
For instance, a test case with \texttt{PASSED}, \texttt{FAILED}, and \texttt{PASSED} statuses in run.log, test.log, and fix.log, respectively, is represented as \texttt{PASSED}→\texttt{FAILED}→\texttt{PASSED}.
We apply the following filtering rules to determine eligible PRs:
\begin{itemize}[leftmargin=*]
    \item PRs with any \texttt{ANY}→\texttt{PASSED}→\texttt{FAILED} transitions are discarded to ensure that no potential regressions are introduced by the fix.patch.
    \item PRs without at least one \texttt{ANY}→\texttt{FAILED}→\texttt{PASSED} transition are discarded, as they do not demonstrate any effective bug fix.
    \item PRs exhibiting abnormal transitions such as \texttt{PASSED}→\texttt{NONE}/\texttt{SKIPPED}→\texttt{FAILED} are discarded to eliminate ambiguous test behaviors.
\end{itemize}
After applying these criteria, we retain $2,456$ issue-resolving instances spanning $39$ repositories across $7$ languages.
For each instance, we extract test cases exhibiting transitions of the form \texttt{Any}→\texttt{FAILED}/\texttt{PASSED}/\texttt{SKIPPED}/\texttt{NONE}→\texttt{PASSED}, and include them in the dataset to enable fine-grained and reliable evaluation.

\begin{table*}[!ht]
    \centering
    \caption{Statistics of the \multiswebench.
    \#A2P2P, \#A2F2P, and \#A2N2P represent the average counts of \texttt{Any}→\texttt{PASSED}\&\texttt{FAILED}\&\texttt{NONE}→\texttt{PASSED} unit tests.}
    \vspace{-2mm}
    \scalebox{0.62}{
    \begin{tabular}{lcc|c|c|ccc|ccc}
    \toprule
     \multicolumn{3}{c|}{\textbf{Repository}}& \multicolumn{1}{c|}{\textbf{Instance}} & \multicolumn{1}{c|}{\textbf{Issue description}} & \multicolumn{3}{c|}{\textbf{Fix patches}} & \multicolumn{3}{c}{\textbf{Unit tests}} \\
    \hline
    Org/Repo & \#Files & \#LoC & \#Num & Avg. \#Tokens & Avg. \#Lines & Avg. \#Hunks & Avg. \#Files & \#A2P2P & \#A2F2P & \#A2N2P \\
    \hline
    \rowcolor{mygray} \multicolumn{11}{c}{Java} \\
\href{https://github.com/alibaba/fastjson2}{alibaba/fastjson2} & 4244 & 443.8k & 6 & 459.2 & 10.5 & 1.3 & 1.2 & 1243.5 & 0.8 & 1020.5 \\ 
\href{https://github.com/elastic/logstash}{elastic/logstash} & 562 & 59.9k & 38 & 1600.4 & 212.3 & 10.0 & 4.6 & 554.7 & 1.9 & 256.2 \\ 
\href{https://github.com/mockito/mockito}{mockito/mockito} & 986 & 84.0k & 6 & 315.2 & 92.5 & 10.3 & 4.7 & 97.2 & 1.0 & 3.8 \\ 
\href{https://github.com/apache/dubbo}{apache/dubbo} & 3939 & 402.1k & 3 & 774.0 & 9.3 & 3.0 & 1.3 & 2.0 & 57.0 & 0.0 \\ 
\href{https://github.com/fasterxml/jackson-core}{fasterxml/j-core} & 366 & 105.7k & 18 & 304.7 & 33.8 & 4.8 & 2.1 & 2.0 & 85.6 & 0.0 \\ 
\href{https://github.com/fasterxml/jackson-databind}{fasterxml/j-dbind} & 1230 & 217.5k & 42 & 621.5 & 35.1 & 3.9 & 2.1 & 2.0 & 73.8 & 0.0 \\ 
\href{https://github.com/fasterxml/jackson-dataformat-xml}{fasterxml/j-dfmt-xml} & 206 & 23.0k & 5 & 1071.8 & 98.4 & 10.4 & 3.2 & 2.0 & 94.2 & 0.0 \\ 
\href{https://github.com/google/gson}{google/gson} & 261 & 48.0k & 5 & 365.8 & 35.8 & 4.6 & 1.8 & 2.0 & 62.6 & 0.0 \\ 
\href{https://github.com/googlecontainertools/jib}{google-ct/jib} & 604 & 75.5k & 5 & 1094.6 & 15.2 & 3.2 & 2.6 & 2.0 & 96.2 & 0.0 \\ 
\hline
    \rowcolor{mygray} \multicolumn{11}{c}{TypeScript} \\
\href{https://github.com/darkreader/darkreader}{darkreader/darkreader} & 189 & 26.2k & 2 & 749.5 & 13.0 & 2.0 & 1.5 & 41.0 & 3.5 & 0.0 \\ 
\href{https://github.com/mui/material-ui}{mui/material-ui} & 27632 & 698.6k & 174 & 508.6 & 331.2 & 20.2 & 12.0 & 5001.3 & 2.3 & 836.8 \\ 
\href{https://github.com/vuejs/core}{vuejs/core} & 509 & 128.2k & 48 & 694.8 & 22.9 & 3.5 & 1.9 & 2920.4 & 3.0 & 0.0 \\ 
    \hline
    \rowcolor{mygray} \multicolumn{11}{c}{JavaScript} \\
\href{https://github.com/anuraghazra/github-readme-stats}{ag/gh-rdme-stats} & 69 & 11.8k & 19 & 287.1 & 123.6 & 13.5 & 4.8 & 108.9 & 3.5 & 3.4 \\ 
\href{https://github.com/axios/axios}{axios/axios} & 166 & 21.0k & 4 & 490.8 & 179.5 & 7.8 & 4.0 & 68.5 & 1.2 & 0.0 \\ 
\href{https://github.com/expressjs/express}{expressjs/express} & 142 & 17.3k & 4 & 177.5 & 7.2 & 2.2 & 1.5 & 808.2 & 1.5 & 65.2 \\ 
\href{https://github.com/iamkun/dayjs}{iamkun/dayjs} & 324 & 17.1k & 56 & 325.6 & 21.7 & 2.7 & 2.0 & 60.4 & 1.2 & 3.2 \\ 
\href{https://github.com/Kong/insomnia}{Kong/insomnia} & 526 & 182.0k & 1 & 709.0 & 1.0 & 1.0 & 1.0 & 105.0 & 1.0 & 0.0 \\ 
\href{https://github.com/sveltejs/svelte}{sveltejs/svelte} & 2800 & 105.9k & 272 & 618.9 & 72.0 & 8.4 & 4.0 & 4904.2 & 5.5 & 0.0 \\ 
    \hline
    \rowcolor{mygray} \multicolumn{11}{c}{Go} \\
\href{https://github.com/cli/cli}{cli/cli} & 737 & 165.1k & 397 & 347.6 & 103.8 & 9.0 & 3.9 & 1997.0 & 2.9 & 31.0 \\ 
\href{https://github.com/grpc/grpc-go}{grpc/grpc-go} & 981 & 260.8k & 16 & 276.1 & 81.8 & 7.7 & 2.8 & 230.4 & 0.6 & 6.6 \\ 
\href{https://github.com/zeromicro/go-zero}{zeromicro/go-zero} & 960 & 117.6k & 15 & 205.2 & 52.4 & 4.9 & 2.7 & 1318.9 & 0.3 & 43.9 \\ 
    \hline
    \rowcolor{mygray} \multicolumn{11}{c}{Rust} \\
\href{https://github.com/BurntSushi/ripgrep}{BurntSushi/ripgrep} & 98 & 45.4k & 14 & 553.7 & 1604.9 & 21.9 & 7.5 & 233.2 & 1.1 & 8.1 \\ 
\href{https://github.com/clap-rs/clap}{clap-rs/clap} & 321 & 70.4k & 132 & 987.0 & 147.1 & 15.7 & 4.7 & 489.5 & 3.1 & 378.8 \\ 
\href{https://github.com/nushell/nushell}{nushell/nushell} & 1479 & 264.2k & 14 & 795.6 & 155.0 & 10.6 & 4.3 & 798.6 & 2.6 & 336.6 \\ 
\href{https://github.com/rayon-rs/rayon}{rayon-rs/rayon} & 191 & 36.9k & 2 & 153.5 & 637.5 & 5.5 & 2.0 & 113.5 & 0.5 & 171.0 \\ 
\href{https://github.com/serde-rs/serde}{serde-rs/serde} & 188 & 36.5k & 2 & 171.5 & 72.5 & 3.0 & 3.0 & 0.0 & 0.0 & 294.5 \\ 
\href{https://github.com/sharkdp/bat}{sharkdp/bat} & 83 & 22.0k & 10 & 638.2 & 239.5 & 14.1 & 5.9 & 152.7 & 1.7 & 33.6 \\ 
\href{https://github.com/sharkdp/fd}{sharkdp/fd} & 24 & 6.7k & 14 & 167.8 & 55.8 & 7.8 & 4.5 & 186.5 & 1.1 & 0.0 \\ 
\href{https://github.com/tokio-rs/bytes}{tokio-rs/bytes} & 33 & 11.9k & 5 & 188.0 & 45.0 & 5.6 & 1.8 & 23.2 & 0.4 & 91.6 \\ 
\href{https://github.com/tokio-rs/tokio}{tokio-rs/tokio} & 727 & 141.5k & 25 & 590.0 & 139.8 & 10.6 & 3.5 & 26.6 & 0.0 & 287.4 \\ 
\href{https://github.com/tokio-rs/tracing}{tokio-rs/tracing} & 241 & 60.9k & 21 & 472.0 & 597.2 & 39.3 & 7.1 & 30.8 & 0.2 & 182.0 \\ 
\hline
    \rowcolor{mygray} \multicolumn{11}{c}{C} \\
\href{https://github.com/facebook/zstd}{facebook/zstd} & 276 & 119.8k & 29 & 496.6 & 67.6 & 10.9 & 3.0 & 0.8 & 0.5 & 5.6 \\ 
\href{https://github.com/jqlang/jq}{jqlang/jq} & 80 & 43.0k & 17 & 429.8 & 26.1 & 2.7 & 1.8 & 27.2 & 1.0 & 0.1 \\ 
\href{https://github.com/ponylang/ponyc}{ponylang/ponyc} & 285 & 80.2k & 82 & 480.2 & 205.4 & 15.6 & 5.7 & 997.6 & 1.9 & 388.8 \\ 
    \hline
    \rowcolor{mygray} \multicolumn{11}{c}{C++} \\
\href{https://github.com/catchorg/Catch2}{catchorg/Catch2} & 399 & 58.0k & 12 & 357.3 & 469.0 & 15.4 & 8.2 & 19.9 & 0.7 & 17.6 \\ 
\href{https://github.com/fmtlib/fmt}{fmtlib/fmt} & 25 & 36.4k & 41 & 397.7 & 36.8 & 3.0 & 1.1 & 9.3 & 0.0 & 9.3 \\ 
\href{https://github.com/nlohmann/json}{nlohmann/json} & 477 & 124.7k & 55 & 905.5 & 405.8 & 27.9 & 6.5 & 26.5 & 0.0 & 42.9 \\ 
\href{https://github.com/simdjson/simdjson}{simdjson/simdjson} & 455 & 229.7k & 20 & 320.2 & 768.5 & 35.5 & 11.0 & 18.6 & 0.0 & 41.5 \\ 
\href{https://github.com/yhirose/cpp-httplib}{yhirose/cpp-httplib} & 33 & 50.9k & 1 & 240.0 & 1.0 & 1.0 & 1.0 & 272.0 & 1.0 & 0.0 \\ 
    \bottomrule
    \end{tabular}}
    \label{tab:repo_stat}
\end{table*}

\subsubsection{Phase 5: Manual Verification}
\label{sec:phase5_manual_verification}
To ensure the reliability of \multiswebench in evaluating the issue-resolving capabilities of LLMs, we conduct comprehensive manual verification on the $2,456$ issue-resolving instances retained from the previous phase.
Our verification process follows the annotation guidelines of the recently released SWE-bench-verified\footnote{\url{https://openai.com/index/introducing-swe-bench-verified}}.
In detail, we recruit $68$ annotators through outsourcing, with the number per language proportional to the remaining annotation workload.
All annotators were screened based on their qualifications, including at least two years of experience in the target language and a relevant bachelor's degree or higher.

Before annotation, each annotator undergoes a one-hour training session covering the background, objectives, procedures, deliverables, and quality standards of the task.
To further support consistency and correctness during the annotation process, we establish dedicated discussion channels to provide real-time guidance and handle edge cases collaboratively.
Each instance is independently labeled by two annotators. 
Upon completion, the two annotations are cross-reviewed to produce a single, agreed-upon final label.
To ensure high annotation quality, we additionally form an internal quality assessment team of $14$ experienced engineers.
This team produces reference answers and verifies that the outsourced annotations for each language reach a minimum accuracy threshold of $80\%$.
After rigorous manual verification, a total of $1,632$ high-quality instances are retained as the final \multiswebench, filtered by annotation criteria from the verification questionnaire\footnote{\url{https://github.com/multi-swe-bench/multi-swe-bench/blob/main/docs/manual-verification/questionnaire-demo.pdf}}: Q2.1=0 \& Q3.1$\in$\{2,3\} \& Q4.1$\in$\{2,3\}.
All annotation results have been made publicly available to ensure the transparency of the dataset\footnote{\url{https://github.com/multi-swe-bench/multi-swe-bench/tree/main/docs/manual-verification/annotation-results}}.

\subsection{Features of \multiswebench}
\label{sec:features_of_multiswebench}
Tab.~\ref{tab:repo_stat} presents an overview of the key statistics of \multiswebench, highlighting its coverage across a wide range of programming languages and repositories.
It includes $1,632$ issue-resolving instances sourced from $39$ diverse repositories, spanning $7$ popular languages: Java, TypeScript, JavaScript, Go, Rust, C, and C++.
These repositories vary significantly in size and complexity, with the number of files ranging from $24$ to $27,632$, and lines of code from $6.7$k to $698.6$k. 
This diversity ensures that \multiswebench reflects realistic and heterogeneous software development scenarios.
In terms of issue descriptions, the complexity varies notably across repositories and languages. 
Java and Rust projects generally present longer and more detailed issue reports (e.g., "elastic/logstash" with $1600.4$ tokens), suggesting more context-dependent reasoning is required. 
In contrast, JavaScript, Go, and C issues are typically brief and focused (e.g., "expressjs/express" with $177.5$ tokens), implying simpler or more localized fixes.
This variation in description length highlights the need for LLMs to adapt to both under- and over-specified problem statements.
Similarly, patch complexity also differs significantly across languages. 
Rust and C++ projects frequently require large-scale edits, with some instances modifying over $200$ lines and $7$ files per patch (e.g., "BurntSushi/ripgrep" and "simdjson/simdjson"). 
Conversely, JavaScript, and TypeScript patches tend to be more localized and atomic, often involving under $3$ hunks and fewer than $2$ files. 
These contrasts emphasize the importance of handling both high-granularity refactoring and precision editing.
Moreover, all repositories come with strong test coverage, providing reliable signals for verifying patch correctness, as confirmed by the manual verification in Sec.~\ref{sec:phase5_manual_verification}.

\begin{figure}[t]
    \centering
    \includegraphics[width=0.55\linewidth]{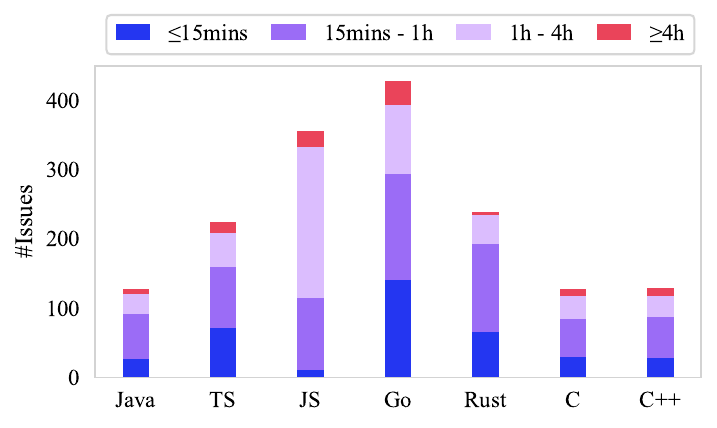}
    \caption{Distribution of estimated time consumption of issues in \multiswebench.}
    \label{fig:enter-label}
    \vspace{-0.5cm}
\end{figure}

We further display the manual verification results in Tab.~\ref{tab:dataset}, which show that most instances have no serious issues and receive high scores, confirming the overall quality of the repositories selected in Sec.~\ref{sec:phase1_repository_selection}.
As part of the manual annotation process in \multiswebench, we recorded the estimated time required to resolve each issue, categorized into four buckets: $\leq$15 minutes, 15 minutes–1 hour, 1–4 hours, and $\geq$4 hours (Fig.~\ref{fig:enter-label}). 
Unlike SWE-Bench, we use this time-based annotation to define difficulty levels across all languages: easy ($\leq$15 mins), medium (15 mins–1h), and hard ($\geq$1h). 
Tab.~\ref{tab:difficult_stat} summarizes the distribution of difficulty levels by language. 
We observe clear trends across these categories: 
As difficulty increases, issues tend to have longer descriptions, and the corresponding patches involve more lines, hunks, and files. 
Interestingly, certain easy instances exhibit large-scale edits (e.g., Rust), which are typically due to highly repetitive and pattern-consistent changes. 
This highlights the advantage of time-based difficulty categorization over superficial metrics like token count or file span. 
Such categorization provides a more realistic measure of problem complexity and can better guide the development and evaluation of LLMs.

\begin{table*}[!ht]
    \centering
    \caption{Feature distribution of \multiswebench instances by difficulty and language.}
    \vspace{-2mm}
    \scalebox{0.7}{
    \begin{tabular}{l|c|c|ccc|ccc}
    \toprule
     \multirow{2}{*}{\textbf{Difficulty}} & \multicolumn{1}{c|}{\textbf{Instance}} & \multicolumn{1}{c|}{\textbf{Issue description}} & \multicolumn{3}{c|}{\textbf{Fix patches}} & \multicolumn{3}{c}{\textbf{Unit tests}} \\
    \cline{2-9}  
     & \#Num & Avg. \#Tokens   & Avg. \#Lines & Avg. \#Hunks & Avg. \#Files & \#A2P2P & \#A2F2P & \#A2N2P \\
    \hline
    \rowcolor{mygray} \multicolumn{9}{c}{Python} \\
Easy & 194 & 417.9 & 5.0 & 1.4 & 1.0 & 116.2 & 3.9 & 0\\ 
Medium &  261 & 555.9 & 14.1 & 2.5 & 1.3 & 115.4 & 2.4 & 0\\ 
Hard & 45 & 589.8 & 55.8 & 6.8 & 2.0 & 166.3 & 2.9 & 0\\ 
    \hline
    \rowcolor{mygray} \multicolumn{9}{c}{Java} \\
Easy & 27 & 733.8 & 12.4 & 2.6 & 1.8 & 126.8 & 58.0 & 76.1 \\ 
Medium & 65 & 843.3 & 36.2 & 4.6 & 2.1 & 182.3 & 58.6 & 136.9 \\ 
Hard & 36 & 1039.0 & 246.1 & 11.9 & 5.4 & 389.1 & 21.8 & 136.9 \\ 
    \hline
     \rowcolor{mygray} \multicolumn{9}{c}{TypeScript} \\
Easy & 72 & 600.1 & 8.3 & 2.1 & 1.5 & 4806.8 & 2.0 & 0.0 \\ 
Medium & 88 & 566.9 & 74.3 & 8.8 & 4.3 & 4854.6 & 2.8 & 214.3 \\ 
Hard & 64 & 472.8 & 806.6 & 43.2 & 26.5 & 3706.1 & 2.7 & 1980.4 \\ 
    \hline
    \rowcolor{mygray} \multicolumn{9}{c}{JavaScript} \\
Easy & 10 & 282.4 & 4.7 & 1.8 & 1.6 & 616.8 & 1.2 & 35.1 \\ 
Medium & 105 & 505.8 & 15.5 & 2.6 & 2.1 & 3161.0 & 3.6 & 0.8 \\ 
Hard & 241 & 578.7 & 92.2 & 10.1 & 4.5 & 4169.9 & 5.2 & 0.3 \\ 
    \hline
    \rowcolor{mygray} \multicolumn{9}{c}{Go} \\
Easy & 141 & 411.7 & 26.6 & 4.0 & 2.7 & 2181.0 & 2.6 & 20.4 \\ 
Medium & 153 & 331.4 & 49.6 & 6.9 & 2.6 & 1832.5 & 2.2 & 25.7 \\ 
Hard & 134 & 274.0 & 238.6 & 16.0 & 6.6 & 1704.2 & 3.4 & 46.7 \\ 
    \hline
    \rowcolor{mygray} \multicolumn{9}{c}{Rust} \\
Easy & 66 & 808.2 & 318.7 & 7.0 & 3.3 & 465.2 & 3.2 & 212.0 \\ 
Medium & 126 & 814.7 & 113.6 & 10.6 & 3.7 & 343.0 & 1.8 & 300.5 \\ 
Hard & 47 & 599.4 & 629.0 & 45.2 & 10.3 & 232.3 & 1.1 & 334.0 \\ 
    \hline
    \rowcolor{mygray} \multicolumn{9}{c}{C} \\
Easy & 30 & 551.4 & 16.4 & 3.7 & 2.2 & 424.8 & 0.8 & 208.2 \\ 
Medium & 54 & 449.9 & 36.7 & 5.5 & 2.5 & 715.5 & 1.0 & 228.2 \\ 
Hard & 44 & 460.2 & 381.1 & 28.0 & 8.7 & 702.5 & 2.4 & 306.3 \\ 
    \hline
    \rowcolor{mygray} \multicolumn{9}{c}{C++} \\
Easy & 28 & 494.5 & 25.2 & 4.4 & 2.2 & 45.0 & 0.1 & 15.7 \\ 
Medium & 59 & 427.5 & 204.2 & 7.6 & 3.3 & 18.2 & 0.1 & 23.0 \\ 
Hard & 42 & 904.2 & 763.7 & 47.2 & 11.1 & 9.3 & 0.0 & 47.2 \\  
    \bottomrule
    \end{tabular}}
    \label{tab:difficult_stat}
\end{table*}

\begin{table*}[!ht]
    \centering
    \caption{Scoring statistics for \multiswebench from the verification questionnaire.}
    \scalebox{0.7}{
    \begin{tabular}{l|cc|cccc|cccc}
    \toprule
     \multirow{2}{*}{\textbf{Languages}} & \multicolumn{2}{c|}{\textbf{Q2.1 Serious Issue Flag}} & \multicolumn{4}{c|}{\textbf{Q3.1 Clarity of Issue Description}} & \multicolumn{4}{c}{\textbf{Q4.1 Coverage of Unit Tests}} \\
      \cline{2-11}
       & \#Score 0 & \#Score 1 & \#Score 0 & \#Score 1 & \#Score 2 & \#Score 3 &\#Score 0 & \#Score 1 & \#Score 2 & \#Score 3 \\
    \hline
Java & 146 & 10 & 2 & 2 & 44 & 98 & 10 & 5 & 17 & 114 \\ 
TypeScript & 382 & 8 & 5 & 56 & 121 & 200 & 31 & 76 & 133 & 142 \\ 
JavaScript & 586 & 4 & 0 & 6 & 13 & 567 & 55 & 172 & 305 & 54 \\ 
Go & 579 & 26 & 5 & 10 & 276 & 288 & 44 & 100 & 151 & 284 \\ 
Rust & 328 & 11 & 4 & 20 & 165 & 139 & 23 & 50 & 74 & 181 \\ 
C & 200 & 6 & 2 & 4 & 115 & 79 & 13 & 55 & 83 & 49 \\ 
C++ & 162 & 7 & 0 & 6 & 96 & 60 & 7 & 21 & 45 & 89 \\ 
    \bottomrule
    \end{tabular}
}
    \label{tab:dataset}
\end{table*}

\section{\multiswerl Open-Source Community}
\label{sec:multiswerl_opensource_community}

\noindent
\textbf{Community Introduction.}
\multiswerl is an open-source community aimed at developing high-quality RL training datasets for complex software engineering tasks.
Its purpose is to serve as the foundational infrastructure for training fully autonomous agents capable of addressing real-world software engineering challenges, paving the way toward achieving AGI.
The need for such a community has become increasingly urgent as the potential of RL continues to expand.
Notable models such as DeepSeek-R1~\citep{deepseekr1}, OpenAI o1~\citep{openaio1}, and o3~\citep{openai_o3_mini} have demonstrated the power of RL, even with simple, rule-based reward signals.
In light of these advancements, we are firmly convinced that “\textit{scaling RL in real-world environments is the path toward human-like intelligence}”.
However, the creation of such interactive environments and data trajectories is extremely challenging.
For instance, the development of our \multiswebench took about one year to produce just high-quality $1,632$ instances.
Therefore, we launched the \multiswerl community to harness the power of open-source collaborative contributions for building diverse RL environments.

\noindent
\textbf{Community Initialization.}
To bootstrap the \multiswerl community, we release an initial dataset comprising $4,723$ issue-resolving instances spanning $76$ widely-used open-source repositories and $7$ programming languages: Java, TypeScript, JavaScript, Go, Rust, C, and C++.
Each instance is equipped with a fully containerized execution environment to ensure reproducibility and ease of integration.
This dataset was constructed using the same pipeline as \multiswebench, excluding the manual verification process described in Sec.~\ref{sec:phase5_manual_verification}.
Details about this release are available at 
\href{https://huggingface.co/datasets/ByteDance-Seed/Multi-SWE-RL}{Hugging Face dataset} and \href{https://docs.google.com/spreadsheets/d/1C90SiRmlac3FizmsJzxzrhSNsnCjyYewdrXzFbBV4x0/edit?gid=493937140#gid=493937140}{\multiswerl contribution board}.
We envision this initial release as a spark—igniting broader community collaboration and fueling the construction of scalable, high-quality RL environments for real-world software engineering.

\noindent
\textbf{Contribution Guidelines and Recognition.}
We welcome contributions from the community to expand the \multiswebench and \multiswerl.
To help new contributors get started, we provide a detailed demo that walks through the process of creating an issue-resolving instance, available at
\href{https://github.com/multi-swe-bench/multi-swe-bench/blob/main/docs/contribution-demo.md}{Contribution-demo.md}.
To recognize and incentivize community contributions, we maintain a rolling update schedule through periodic arXiv updates or follow-up technical reports, with new versions released every three months.
Each update may include:
\begin{itemize}
  \item Newly added benchmarks for additional programming languages in \multiswebench, with new authors and contributors;
  \item Newly contributed data to \multiswerl, with new authors and contributors;
  \item Newly reported performance results from RL trials on \multiswebench using \multiswerl data, with new authors and contributors;
  \item Newly open-sourced RL models with significantly enhanced performance, with new authors and contributors.
\end{itemize}
Our contribution incentive policy is detailed at \href{https://github.com/multi-swe-bench/multi-swe-bench/blob/main/docs/contribution-incentive-plan.md}{Incentive-plan.md}.
We are committed to continuously refining our contribution strategy to encourage sustained open-source engagement, and we warmly invite the community to take part in shaping and scaling this collaborative effort.

\section{Experimental Setups}

\subsection{Evaluated LLMs and Methods}
\noindent\textbf{Methods.}
We evaluate three representative methods for issue resolving: \agentless~\citep{xia2024agentless}, \sweagent~\citep{yang2024sweagent}, and \openhands+ CodeAct v2.1~\citep{openhands}.
These methods were specifically designed for Python as used in SWE-Bench~\citep{swe-bench}.
We extended the methods to accommodate the multilingual nature of \multiswebench\footnote{\magentless and \mopenhands are pronounced as \textipa{/"mA:dZ@nt.l@s/} and \textipa{/"moUp@n.hAndz/}, respectively.}.
\begin{itemize}[leftmargin=*]
    \item \textbf{\agentless\footnote{\url{https://github.com/OpenAutoCoder/Agentless}}→\magentless\footnote{\url{https://github.com/multi-swe-bench/MagentLess}}}: 
    \agentless addresses the issue resolving task through a multi-stage fixed workflow, including hierarchical fault localization, code repair, and candidate patch selection via regression and reproduction tests.
    In \magentless, we made the following key modifications to support multilingual adaptation and improve scalability:
    \begin{enumerate}
    \item We revised all prompts to accommodate the newly added languages.
    \item We replaced all file skeleton inputs with full file content, as extracting file skeletons is challenging in some programming languages.
    \item We implemented function and class extraction for all languages using Tree-sitter\footnote{\url{https://tree-sitter.github.io}}.
    \item We pruned the extracted repository structures by retaining only files and directories with specific extensions, as repositories in certain languages (e.g., TypeScript) often contain an excessive number of files that may exceed LLM context limits.
    \item We removed the candidate patch selection stage and retained only fault localization and code repair, as regression and reproduction testing is cumbersome to implement across languages and falls outside the scope of this work.
    \end{enumerate}   
    \item \textbf{\sweagent\footnote{\url{https://github.com/SWE-agent/SWE-agent}}→\msweagent\footnote{\url{https://github.com/multi-swe-bench/MSWE-agent}}}: 
    \sweagent is an agent-based approach that solves issues through multi-turn interactions via a predefined agent-computer interface (ACI).
    To support \multiswebench, we developed \msweagent with the following modifications:
    \begin{enumerate}
    \item We revised all prompts to accommodate the newly added languages.
    \item We truncated overly long environment observations to ensure stable agent execution.
    \item We added "\texttt{.gitignore}" to exclude compiled artifacts (e.g., "\texttt{.o}", "\texttt{.bin}") in languages like C/C++, which could otherwise interfere with "\texttt{git apply}".
    \item We fixed language-specific commands that caused crashes or non-terminating behavior during execution to ensure stable agent execution.
\end{enumerate}
    \item \textbf{\openhands\footnote{\url{https://github.com/All-Hands-AI/OpenHands}}→\mopenhands\footnote{\url{https://github.com/multi-swe-bench/MopenHands}}}: 
    \openhands is a widely adopted platform for building software development agents.
    In \mopenhands, we made the following key modifications to support multilingual adaptation:
    \begin{enumerate}
    \item We revised all prompts to support the newly added programming languages.
    \item We added "\texttt{.gitignore}" to exclude compiled artifacts, as also done in \msweagent.
    \item We fixed several implementation bugs, including an issue where "\texttt{CmdRunAction}" incorrectly rendered tab characters (\textbackslash t) as spaces in "\texttt{git diff}" outputs, making patches unapplicable. To resolve this, we redirected the diff to a file and read it using "\texttt{FileReadAction}", which proved especially important in languages like Go.
\end{enumerate}
\end{itemize}
We have systematically extended the above methods to support the multilingual setting of \multiswebench. 
Still, there remains substantial room for improvement, particularly in language-specific adaptation and overall robustness.
We welcome community collaboration to further advance their capabilities.

\noindent\textbf{LLMs.}
We evaluated $9$ popular LLMs across the above three methods: GPT-4o (gpt-4o-2024-11-20), OpenAI-o1 (o1-2024-12-17), OpenAI-o3-mini-high (o3-mini-2025-01-31 high), Claude-3.5-Sonnet (claude-3-5-sonnet-20241022), Claude-3.7-Sonnet (claude-3-7-sonnet-20250219), DeepSeek-V3, DeepSeek-R1, Qwen2.5-72B-Instruct, and Doubao-1.5-pro.

\subsection{Evaluation Metrics}
Following SWE-Bench~\citep{swe-bench} and SWE-Lancer~\citep{swelancer}, we adopt Resolved Rate (\%) as our primary evaluation metric, measuring the percentage of issues resolved.
In addition, we report several other metrics to provide a more detailed analysis: 
Success Location (\%) — the accuracy of fault localization at file level; 
and Average Cost (\$) — the average cost per issue.

\section{Experimental Results}
\subsection{Performance on \multiswebench}

In this subsection, we conduct a systematic evaluation of issue resolving performance on \multiswebench along three key dimensions: 
(1) \emph{language-specific performance}, examining variations in effectiveness across programming languages; 
(2) \emph{LLMs and agents comparison}, evaluating the differential capabilities of various LLMs and methods; 
and (3) \emph{repository-level performance}, analyzing the impact of repository characteristics on resolved rate.

\subsubsection{Performance across Programming Languages}

Tab.~\ref{tab:all_repo_details} presents the overall performance of the agents across eight programming languages, and Tab.~\ref{tab:all_repo_details_all_levels} further details the results across three distinct difficulty levels.
Based on these tables, several key observations can be drawn and outlined below.

\begin{table*}[ht]
    \centering
    \renewcommand{\arraystretch}{1.1}
    \caption{Resolved rate (\%) of different models on \multiswebench.}
    \vspace{-3mm}
    \label{tab:all_repo_details}
    \scalebox{0.8}{
    \begin{tabular}{lcccccccccccccc} 
    \toprule
    \textbf{Models} & \textbf{Python} & \textbf{Java} & \textbf{TS} & \textbf{JS} & \textbf{Go} & \textbf{Rust} & \textbf{C} & \textbf{C++} \\
    \hline
    \rowcolor{mygray}\multicolumn{9}{c}{\textbf{\magentless}} \\
    GPT-4o & 36.20 & 11.72 & 2.23 & 1.40 & 2.80 & 5.86 & 1.56 & 6.98 \\
    OpenAI-o1 & 48.20 & 21.09 & 5.80 & 5.06 & 4.44 & 7.11 & 1.56 & 5.43 \\
    OpenAI-o3-mini-high & 46.40 & 5.47 & 0.45 & 2.81 & 3.97 & 7.95 & 3.91 & 1.55 \\
    Claude-3.5-Sonnet & 42.40 & 14.84 & 4.91 & 1.97 & 5.14 & 5.02 & 1.56 & 3.88 \\
    Claude-3.7-Sonnet & 44.60 & 14.06 & 3.57 & 1.97 & 5.84 & 5.44 & 2.34 & 3.10 \\
    DeepSeek-V3 & 41.00 & 7.03 & 6.70 & 3.37 & 5.37 & 5.02 & 3.13 & 1.55 \\
    DeepSeek-R1 & 42.20 & 22.66 & 6.25 & 4.49 & 3.74 & 6.69 & 0.78 & 3.10 \\
    Qwen2.5-72B-Instruct & 26.80 & 10.94 & 4.46 & 0.84 & 1.40 & 2.51 & 0.78 & 0.78 \\
    Doubao-1.5-pro & 26.20 & 5.47 & 2.23 & 1.12 & 2.10 & 4.18 & 0.00 & 0.00 \\
    \midrule
    \rowcolor{mygray}\multicolumn{9}{c}{\textbf{\msweagent}} \\
    GPT-4o & 18.80 & 12.50 & 0.45 & 0.84 & 2.34 & 2.09 & 1.56 & 2.33 \\
    OpenAI-o1 & 28.80 & 21.88 & 4.02 & 4.21 & 4.67 & 4.18 & 3.91 & 3.88 \\
    OpenAI-o3-mini-high & 28.60 & 16.41 & 4.91 & 4.21 & 3.97 & 5.02 & 2.34 & 5.43 \\
    Claude-3.5-Sonnet & 24.80 & 20.31 & 8.04 & 4.21 & 5.84 & 6.69 & 4.69 & 6.98 \\
    Claude-3.7-Sonnet & 45.80 & 23.44 & 11.16 & 4.78 & 5.37 & 6.69 & 8.59 & 11.63 \\
    DeepSeek-V3 & 4.20 & 11.72 & 2.68 & 2.53 & 4.44 & 5.86 & 2.34 & 7.75 \\
    DeepSeek-R1 & 2.00 & 9.38 & 5.80 & 1.40 & 2.10 & 2.09 & 0.78 & 6.20 \\
    Qwen2.5-72B-Instruct & 8.60 & 2.34 & 0.00 & 0.56 & 0.47 & 0.42 & 1.56 & 0.00 \\
    Doubao-1.5-pro & 12.40 & 7.03 & 1.79 & 1.40 & 2.10 & 1.67 & 2.34 & 6.20 \\
    \midrule
    \rowcolor{mygray}\multicolumn{9}{c}{\textbf{\mopenhands}} \\
    GPT-4o & 25.60 & 9.38 & 0.00 & 1.97 & 3.50 & 3.35 & 0.00 & 3.88 \\
    OpenAI-o1 & 16.00 & 3.91 & 0.45 & 3.65 & 3.74 & 2.51 & 3.13 & 3.88 \\
    OpenAI-o3-mini-high & 20.40 & 10.16 & 0.45 & 3.37 & 2.34 & 5.02 & 1.56 & 6.98 \\
    Claude-3.5-Sonnet & 39.00 & 14.84 & 11.61 & 1.97 & 6.78 & 12.13 & 3.13 & 12.40 \\
    Claude-3.7-Sonnet & 52.20 & 21.88 & 2.23 & 5.06 & 7.48 & 15.90 & 8.59 & 14.73 \\
    DeepSeek-V3 & 27.80 & 9.38 & 1.34 & 1.12 & 0.70 & 4.60 & 3.13 & 7.75 \\
    DeepSeek-R1 & 26.00 & 8.59 & 0.45 & 2.53 & 0.00 & 4.60 & 2.34 & 4.65 \\
    Qwen2.5-72B-Instruct & 4.40 & 3.13 & 0.00 & 0.84 & 1.40 & 1.67 & 0.78 & 2.33 \\
    Doubao-1.5-pro & 8.80 & 0.78 & 0.00 & 1.12 & 1.64 & 0.84 & 0.00 & 3.10 \\
    \bottomrule
    \end{tabular}
}
\end{table*}

\begin{table}
    \centering
    \begin{adjustbox}{angle=270}  
    \begin{minipage}{\textheight} 
    \centering
    \renewcommand{\arraystretch}{1.2}
    \caption{Resolved rate (\%) of different models on \multiswebench across various difficulty levels.}\label{tab:all_repo_details_all_levels}
    \vspace{-3mm}
    \scalebox{0.68}{
    \begin{tabular}{l|cccccccc|cccccccc|cccccccc} 
    \toprule
     \multirow{2}{*}{\textbf{Models}} & \multicolumn{8}{|c}{\textbf{Easy}} & \multicolumn{8}{|c}{\textbf{Medium}} & \multicolumn{8}{|c}{\textbf{Hard}} \\
     \cmidrule{2-25}
     & \textbf{Python} & \textbf{Java} & \textbf{TS} & \textbf{JS} & \textbf{Go} & \textbf{Rust} & \textbf{C} & \textbf{C++} & \textbf{Python} & \textbf{Java} & \textbf{TS} & \textbf{JS} & \textbf{Go} & \textbf{Rust} & \textbf{C} & \textbf{C++} & \textbf{Python} & \textbf{Java} & \textbf{TS} & \textbf{JS} & \textbf{Go} & \textbf{Rust} & \textbf{C} & \textbf{C++} \\
    \hline
    \rowcolor{mygray}\multicolumn{25}{c}{\textbf{\magentless}} \\
    GPT-4o & 55.15 & 22.22 & 4.17 & 20.00 & 6.38 & 13.64 & 3.33 & 17.86 & 27.97 & 13.85 & 1.14 & 1.90 & 1.96 & 1.59 & 1.85 & 6.78 & 2.22 & 0.00 & 1.56 & 0.41 & 0.00 & 6.38 & 0.00 & 0.00 \\
    OpenAI-o1 & 68.04 & 40.74 & 11.11 & 20.00 & 6.38 & 16.67 & 3.33 & 14.29 & 40.23 & 24.62 & 4.55 & 9.52 & 5.88 & 3.17 & 1.85 & 5.08 & 8.89 & 0.00 & 1.56 & 2.49 & 0.75 & 4.26 & 0.00 & 0.00 \\
    OpenAI-o3-mini-high & 67.01 & 14.81 & 1.39 & 30.00 & 9.93 & 22.73 & 6.67 & 3.57 & 38.31 & 4.62 & 0.00 & 4.76 & 1.31 & 2.38 & 1.85 & 1.69 & 4.44 & 0.00 & 0.00 & 0.83 & 0.75 & 2.13 & 4.55 & 0.00 \\
    Claude-3.5-Sonnet & 61.86 & 37.04 & 11.11 & 30.00 & 11.35 & 9.09 & 3.33 & 10.71 & 34.10 & 13.85 & 2.27 & 1.90 & 3.27 & 3.17 & 1.85 & 3.39 & 6.67 & 0.00 & 1.56 & 0.83 & 0.75 & 4.26 & 0.00 & 0.00 \\
    Claude-3.7-Sonnet & 64.43 & 33.33 & 5.56 & 20.00 & 13.48 & 10.61 & 3.33 & 7.14 & 35.63 & 13.85 & 3.41 & 3.81 & 3.92 & 3.17 & 1.85 & 3.39 & 11.11 & 0.00 & 1.56 & 0.41 & 0.00 & 4.26 & 2.27 & 0.00 \\
    DeepSeek-V3 & 57.73 & 18.52 & 11.11 & 30.00 & 9.93 & 15.15 & 6.67 & 0.00 & 34.87 & 6.15 & 6.82 & 4.76 & 4.58 & 0.79 & 1.85 & 3.39 & 4.44 & 0.00 & 1.56 & 1.66 & 1.49 & 2.13 & 2.27 & 0.00 \\
    DeepSeek-R1 & 58.76 & 51.85 & 11.11 & 30.00 & 7.80 & 15.15 & 0.00 & 3.57 & 36.02 & 23.08 & 6.82 & 7.62 & 1.96 & 3.97 & 1.85 & 5.08 & 6.67 & 0.00 & 0.00 & 2.07 & 1.49 & 2.13 & 0.00 & 0.00 \\
    Qwen2.5-72B-Instruct & 44.33 & 33.33 & 6.94 & 20.00 & 3.55 & 6.06 & 0.00 & 3.57 & 18.39 & 7.69 & 4.55 & 0.00 & 0.65 & 1.59 & 1.85 & 0.00 & 0.00 & 0.00 & 1.56 & 0.41 & 0.00 & 0.00 & 0.00 & 0.00 \\
    Doubao-1.5-pro & 39.18 & 14.81 & 1.39 & 10.00 & 3.55 & 10.61 & 0.00 & 0.00 & 20.31 & 4.62 & 3.41 & 0.00 & 2.61 & 1.59 & 0.00 & 0.00 & 4.44 & 0.00 & 1.56 & 1.24 & 0.00 & 2.13 & 0.00 & 0.00 \\
    \midrule
    \rowcolor{mygray}\multicolumn{25}{c}{\textbf{\msweagent}} \\
    GPT-4o & 25.77 & 22.22 & 0.00 & 0.00 & 5.67 & 1.52 & 6.67 & 7.14 & 16.09 & 15.38 & 1.14 & 0.95 & 1.31 & 3.17 & 0.00 & 1.69 & 4.44 & 0.00 & 0.00 & 0.83 & 0.00 & 0.00 & 0.00 & 0.00 \\
    OpenAI-o1 & 40.72 & 48.15 & 4.17 & 10.00 & 8.51 & 6.06 & 10.00 & 7.14 & 24.14 & 23.08 & 4.55 & 6.67 & 3.92 & 3.97 & 3.70 & 5.08 & 4.44 & 0.00 & 3.13 & 2.90 & 1.49 & 2.13 & 0.00 & 0.00 \\
    OpenAI-o3-mini-high & 42.78 & 33.33 & 11.11 & 20.00 & 9.22 & 12.12 & 3.33 & 14.29 & 21.46 & 18.46 & 3.41 & 3.81 & 2.61 & 2.38 & 3.70 & 5.08 & 8.89 & 0.00 & 0.00 & 3.73 & 0.00 & 2.13 & 0.00 & 0.00 \\
    Claude-3.5-Sonnet & 28.35 & 48.15 & 15.28 & 0.00 & 13.48 & 13.64 & 13.33 & 17.86 & 25.67 & 20.00 & 5.68 & 7.62 & 2.61 & 4.76 & 3.70 & 6.78 & 4.44 & 0.00 & 3.13 & 2.90 & 1.49 & 2.13 & 0.00 & 0.00 \\
    Claude-3.7-Sonnet & 61.86 & 44.44 & 20.83 & 0.00 & 10.64 & 13.64 & 20.00 & 28.57 & 40.61 & 27.69 & 9.09 & 7.62 & 4.58 & 2.38 & 7.41 & 11.86 & 6.67 & 0.00 & 3.13 & 3.73 & 0.75 & 8.51 & 2.27 & 0.00 \\
    DeepSeek-V3 & 7.22 & 33.33 & 5.56 & 0.00 & 9.22 & 10.61 & 0.00 & 10.71 & 2.68 & 9.23 & 2.27 & 3.81 & 3.27 & 4.76 & 3.70 & 10.17 & 0.00 & 0.00 & 0.00 & 2.07 & 0.75 & 2.13 & 0.00 & 2.38 \\
    DeepSeek-R1 & 2.58 & 14.81 & 9.72 & 10.00 & 6.38 & 4.55 & 3.33 & 17.86 & 1.92 & 12.31 & 6.82 & 1.90 & 0.00 & 0.79 & 0.00 & 5.08 & 0.00 & 0.00 & 0.00 & 0.83 & 0.00 & 2.13 & 0.00 & 0.00 \\
    Qwen2.5-72B-Instruct & 15.46 & 7.41 & 0.00 & 0.00 & 1.42 & 1.52 & 0.00 & 0.00 & 4.98 & 1.54 & 0.00 & 0.95 & 0.00 & 0.00 & 3.70 & 0.00 & 0.00 & 0.00 & 0.00 & 0.41 & 0.00 & 0.00 & 0.00 & 0.00 \\
    Doubao-1.5-pro & 17.53 & 11.11 & 2.78 & 10.00 & 5.67 & 1.52 & 3.33 & 17.86 & 10.73 & 7.69 & 2.27 & 1.90 & 0.65 & 0.79 & 3.70 & 3.39 & 0.00 & 2.78 & 0.00 & 0.83 & 0.00 & 4.26 & 0.00 & 2.38 \\
    \midrule
    \rowcolor{mygray}\multicolumn{25}{c}{\textbf{\mopenhands}} \\
    GPT-4o & 38.66 & 29.63 & 0.00 & 0.00 & 8.51 & 6.06 & 0.00 & 10.71 & 19.54 & 6.15 & 0.00 & 2.86 & 1.96 & 2.38 & 0.00 & 3.39 & 4.44 & 0.00 & 0.00 & 1.66 & 0.00 & 2.13 & 0.00 & 0.00 \\
    OpenAI-o1 & 18.56 & 7.41 & 1.39 & 30.00 & 8.51 & 3.03 & 6.67 & 7.14 & 16.48 & 4.62 & 0.00 & 6.67 & 1.96 & 2.38 & 3.70 & 5.08 & 2.22 & 0.00 & 0.00 & 1.24 & 0.75 & 2.13 & 0.00 & 0.00 \\
    OpenAI-o3-mini-high & 31.44 & 22.22 & 1.39 & 40.00 & 4.96 & 13.64 & 6.67 & 14.29 & 14.56 & 10.77 & 0.00 & 3.81 & 1.96 & 0.79 & 0.00 & 6.78 & 6.67 & 0.00 & 0.00 & 1.66 & 0.00 & 4.26 & 0.00 & 2.38 \\
    Claude-3.5-Sonnet & 48.97 & 25.93 & 18.06 & 10.00 & 12.77 & 24.24 & 6.67 & 32.14 & 36.02 & 18.46 & 13.64 & 3.81 & 5.23 & 7.14 & 3.70 & 10.17 & 13.33 & 0.00 & 1.56 & 0.83 & 2.24 & 8.51 & 0.00 & 2.38 \\
    Claude-3.7-Sonnet & 71.65 & 48.15 & 2.78 & 30.00 & 11.35 & 21.21 & 13.33 & 32.14 & 44.83 & 23.08 & 2.27 & 7.62 & 9.15 & 13.49 & 11.11 & 15.25 & 11.11 & 0.00 & 1.56 & 2.90 & 1.49 & 14.89 & 2.27 & 2.38 \\
    DeepSeek-V3 & 41.24 & 18.52 & 2.78 & 0.00 & 2.13 & 6.06 & 6.67 & 17.86 & 21.46 & 10.77 & 0.00 & 1.90 & 0.00 & 3.97 & 3.70 & 8.47 & 6.67 & 0.00 & 1.56 & 0.83 & 0.00 & 4.26 & 0.00 & 0.00 \\
    DeepSeek-R1 & 41.24 & 14.81 & 1.39 & 10.00 & 0.00 & 13.64 & 6.67 & 14.29 & 19.16 & 10.77 & 0.00 & 2.86 & 0.00 & 1.59 & 1.85 & 3.39 & 0.00 & 0.00 & 0.00 & 2.07 & 0.00 & 0.00 & 0.00 & 0.00 \\
    Qwen2.5-72B-Instruct & 6.70 & 7.41 & 0.00 & 20.00 & 2.13 & 1.52 & 0.00 & 10.71 & 3.45 & 3.08 & 0.00 & 0.95 & 1.31 & 0.79 & 1.85 & 0.00 & 0.00 & 0.00 & 0.00 & 0.00 & 0.75 & 4.26 & 0.00 & 0.00 \\
    Doubao-1.5-pro & 15.46 & 0.00 & 0.00 & 10.00 & 4.96 & 1.52 & 0.00 & 14.29 & 4.98 & 1.54 & 0.00 & 0.95 & 0.00 & 0.00 & 0.00 & 0.00 & 2.22 & 0.00 & 0.00 & 0.83 & 0.00 & 2.13 & 0.00 & 0.00 \\

    \bottomrule
    \end{tabular}
}
        \end{minipage}
    \end{adjustbox}
\end{table}

\textbf{Limited generalization beyond Python.}
From Tab.~\ref{tab:all_repo_details}, it can be observed that existing LLMs and methods demonstrate strong performance in resolving Python issues but struggle to generalize effectively across other languages.
For example, LLMs such as OpenAI-o1 and Claude-3.7-Sonnet achieve high resolved rates for Python but significantly lower effectiveness for most other languages.
This performance disparity can be attributed to three main factors:
(1) \emph{Benchmark difficulty}: \multiswebench is inherently more challenging than SWE-Bench-Verified, with a higher proportion of medium and hard issues (77.1\% for \multiswebench compared to 61.2\% for SWE-Bench-Verified, as calculated from Tab.~\ref{tab:difficult_stat}).
(2) \emph{Method optimization bias}: The three methods are initially optimized for Python, resulting in a performance bias that limits their effectiveness across other languages.
(3) \emph{Language-specific complexity}: Languages like TS and JS feature dynamic typing, asynchronous execution, and diverse runtime behaviors, while languages like C, C++, Rust, and Go involve manual memory management, complex build systems, and intricate type systems, which add to the difficulty for issue resolving.

\textbf{Performance variations across language domains.}
Languages in \multiswebench can be largely categorized into four domains: high-level general-purpose programming (Python, Java), web development (TS, JS), systems programming (Go, Rust), and low-level/high-performance computing (C, C++).
Based on Tab.~\ref{tab:all_repo_details}, the performance generally follows a hierarchy, with high-level general-purpose languages outperforming systems programming and low-level/high-performance computing languages, while web development languages perform the worst.
Java ranks second after Python, though with a noticeable gap.
Go and Rust exhibit inconsistent performance across models, generally outperforming TS and JS but falling behind Java.
C and C++ show even greater variability, with some LLMs (e.g., Doubao-1.5-pro and Qwen2.5-72B-Instruct) struggling to handle them effectively due to the challenges posed by manual memory management and complex compilation pipelines.
TS and JS consistently yield the lowest resolved rates, highlighting the difficulty LLMs face in handling their event-driven, asynchronous programming paradigms.

\textbf{High sensitivity to issue difficulty.}
As shown in Tab.~\ref{tab:all_repo_details_all_levels}, LLM-based agents exhibit a performance that closely aligns with human-labeled difficulty, with resolved rates significantly decreasing as the issue difficulty increases from easy to hard.
However, there are several exceptions. 
For example, the JS language on the \sweagent and Claude-3.5-Sonnet exhibits a zero resolved rate for easy tasks, compared to 7.62\% for medium tasks. 
This suggests that the \msweagent is less effective for JS. 
In addition to human-assigned difficulty, other factors, such as the number of files requiring modification to resolve the issue, also influence performance.
For hard-level issues, existing LLMs and agents are mostly ineffective, with resolved rates approaching zero. 
This phenomenon indicates the limitations of these LLMs and agents: they are primarily capable of addressing issues that human developers can resolve in under 15 minutes and are insufficient for handling more complex tasks requiring over one hour of human effort. 
This finding further underscores the need for RL techniques aimed at advancing agents towards more human-like intelligence, particularly for tackling real-world complex scenarios.

\subsubsection{Performance across Various Methods and LLMs}
\begin{figure}[!t]
    \centering
    \includegraphics[width=0.95\linewidth]{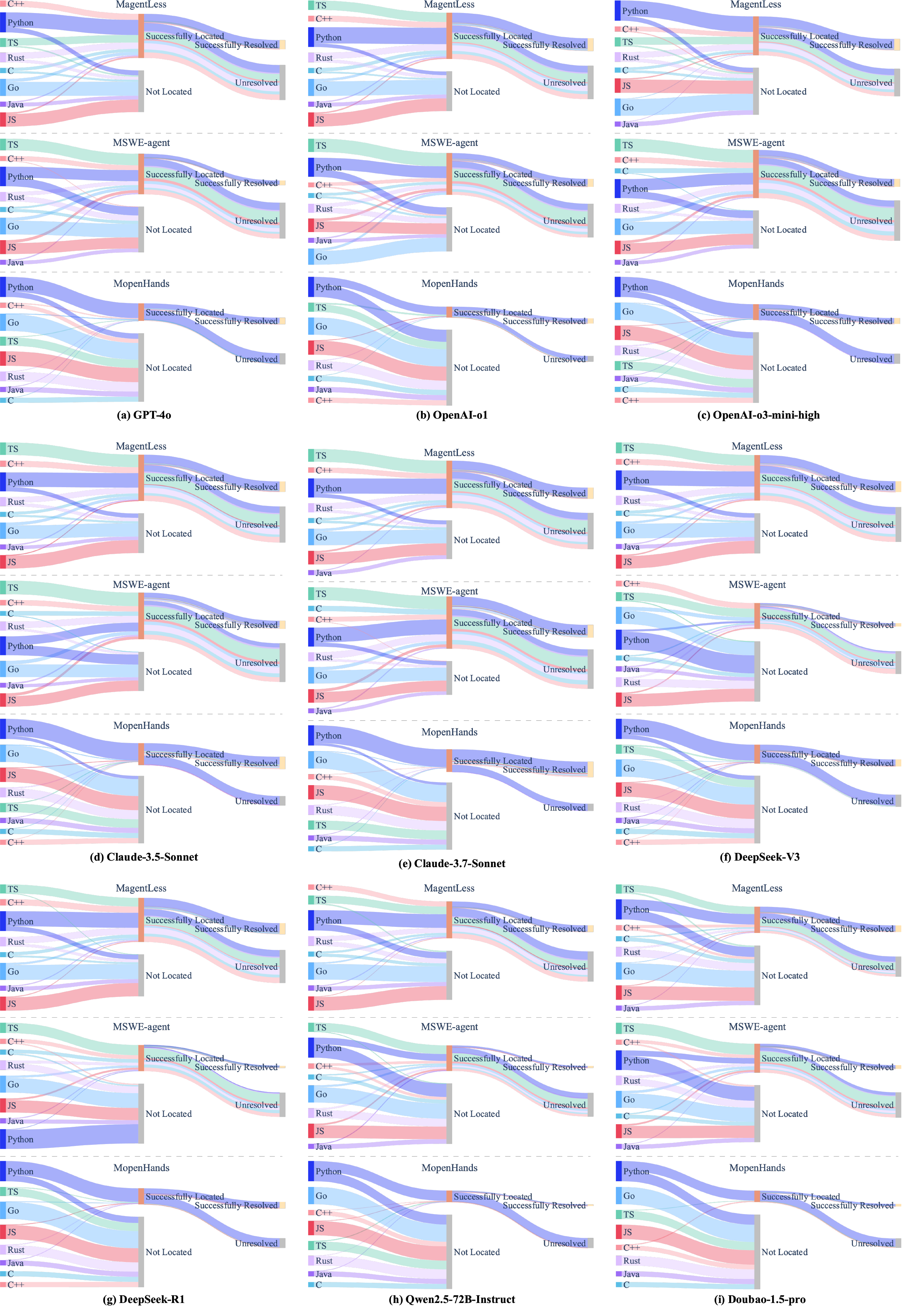}
    \caption{Issue flow from locating to resolving.}
    \label{fig:issue_flow}
\end{figure}

\textbf{Variation in LLMs' performance.}
From Tab.~\ref{tab:all_repo_details}, significant variability is observed in the performance of different LLMs across programming languages.
Specifically, LLMs such as OpenAI-o1, OpenAI-o3-mini-high, Claude-3.5-Sonnet, and Claude-3.7-Sonnet show relatively strong performance, particularly in languages like Python, Java, and C. 
In contrast, LLMs like Qwen2.5-72B-Instruct and Doubao-1.5-pro exhibit rather lower resolved rates, particularly for hard-level issues shown in Tab.~\ref{tab:all_repo_details_all_levels}.
Furthermore, the LLMs exhibit distinct language-specific biases in their performance. 
For example, models like OpenAI-o1 and OpenAI-o3-mini-high perform consistently well across languages such as Python and Java, whereas they struggle significantly with languages like C and C++. 
On the one hand, this performance disparity is likely attributed to the models being better suited to handle higher-level languages like Python and Java, which are more prevalent in their training data. 
On the other hand, the challenges with C and C++ may arise from the models' limited exposure to low-level language features, such as memory management and pointer manipulation, which are less represented in the training data.

\textbf{Performance comparison of methods.}
Tab.~\ref{tab:all_repo_details} and Tab.~\ref{tab:all_repo_details_all_levels} also provide a comprehensive comparison of the resolved rates across three issue-resolving methods, including \magentless, \msweagent, and \mopenhands. 
Overall, \mopenhands outperforms the others in most cases, achieving the highest resolved rate in five out of seven languages, while \msweagent wins twice and \magentless wins once.
The better performance of \mopenhands and \msweagent can be attributed to their more flexible workflow, which is better suited to another language beyond Python compared to \magentless.
In contrast, \magentless follows a more rigid workflow optimized for Python, and the adaptation made to create the \magentless limits its adaptability across other languages.
However, a notable exception to this general trend is observed in the performance of the models DeepSeek-R1 and Qwen2.5-72B-Instruct. 
For these two models, \magentless generally provides better results than \msweagent for languages except C and C++. This suggests that these models may be better suited to the fixed workflow of \magentless.

\textbf{Prioritizing accurate locating over editing and reproducing.}
\magentless, \msweagent, and \mopenhands generally resolve issues through two key steps: issue location and code editing to resolve the issue. 
To provide a more detailed analysis of how existing LLMs and methods perform across these steps, we present the issue flow in Fig.~\ref{fig:issue_flow}.
An issue is considered successfully located if the fix patches generated by the LLMs hit the files of ground truth fix patches.
As shown in Fig.~\ref{fig:issue_flow}, all three methods generally fail to locate issues more often than they succeed.
Accurate issue localization is fundamental to the overall success of the resolution process, serving as a prerequisite for effective code editing. 
Compared to MopenHands, MagentLess achieves more accurate issue localization but struggles more with the code editing step, leading to a lower overall resolved rate. This disparity is particularly evident on Claude-3.7-Sonnet.
This underscores the need for a balanced method that not only prioritizes precise issue identification but also enhances the model’s ability to generate effective fixes. 

\begin{figure}[t]
    \centering
    \includegraphics[width=1\linewidth]{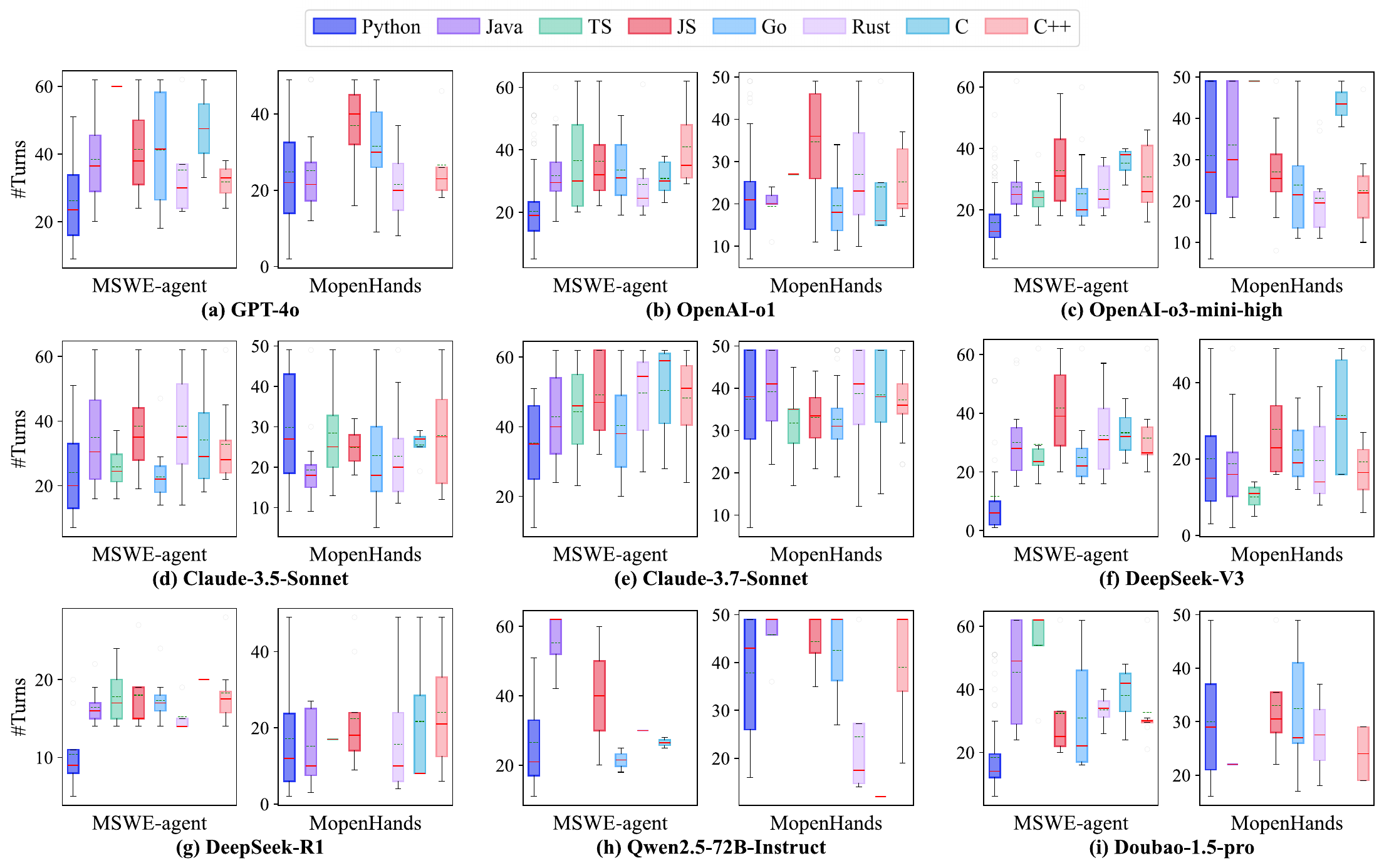}
    \caption{Number of turns required across different programming languages.}
    \label{fig:iterations}
\end{figure}

\textbf{Number of turns required by \msweagent and \mopenhands.}
Both \msweagent and \mopenhands resolve the issue by multi-turn interactions. 
Fig.~\ref{fig:iterations} shows the distribution of turns for successfully resolved an issue.
The absence of a corresponding box plot indicates cases where no issues were successfully resolved, such as \msweagent with Qwen2.5-72B-Instruct on C++.
The number of interaction turns required by two methods differs across models and languages. 
Specifically, \mopenhands resolves issues in fewer turns than \msweagent when using GPT-4o for Java, whereas \msweagent requires fewer turns when resolving Python issues.
However, \mopenhands exhibits a rather higher degree of dispersion in the number of interaction turns compared to \msweagent, which is particularly evident on OpenAI-o3-mini-high.
This suggests that \mopenhands' performance is less stable across different issues, requiring a varying number of turns depending on the complexity or nature of the issue.

\subsubsection{Performance across Different Repositories}

To understand how repository characteristics affect performance, we examine two factors: (1) repository quality, which includes the number of stars, forks, PRs, and issues, and (2) repository complexity, which includes the number of code lines and files, and the language entropy.

\begin{figure}[t]
    \centering
    \includegraphics[width=1\linewidth]{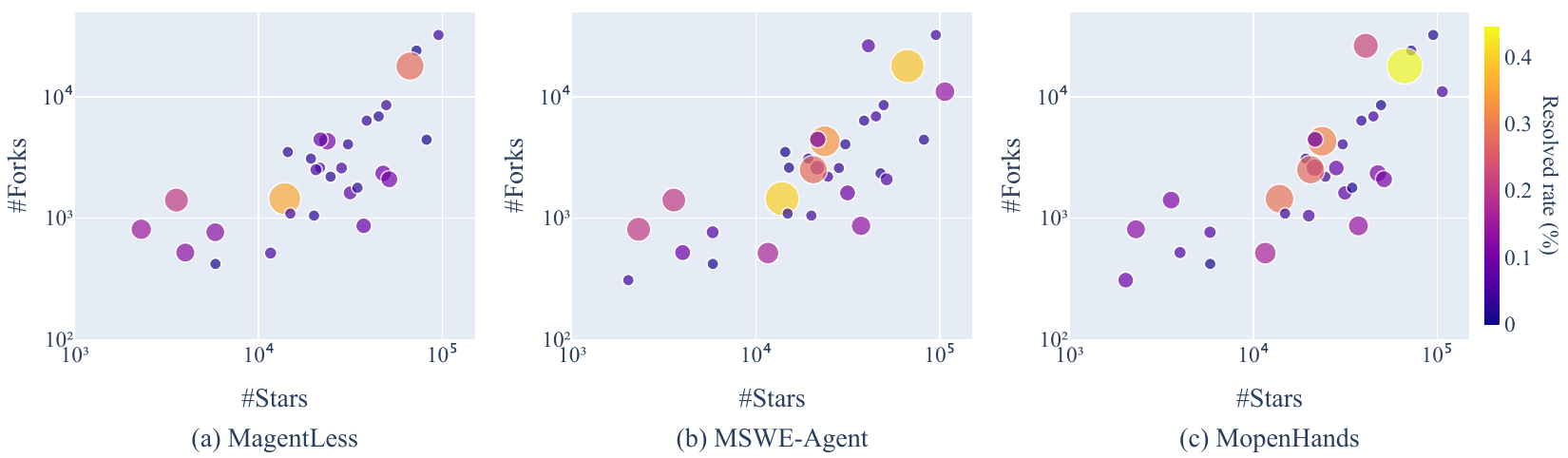}
    \vspace{-3mm}
    \caption{Relationship between resolved rate and the number of stars and forks of a repository.}
    \label{fig:relation_with_repo_quality}
\end{figure}
\begin{figure}[t]
    \centering
    \includegraphics[width=1\linewidth]{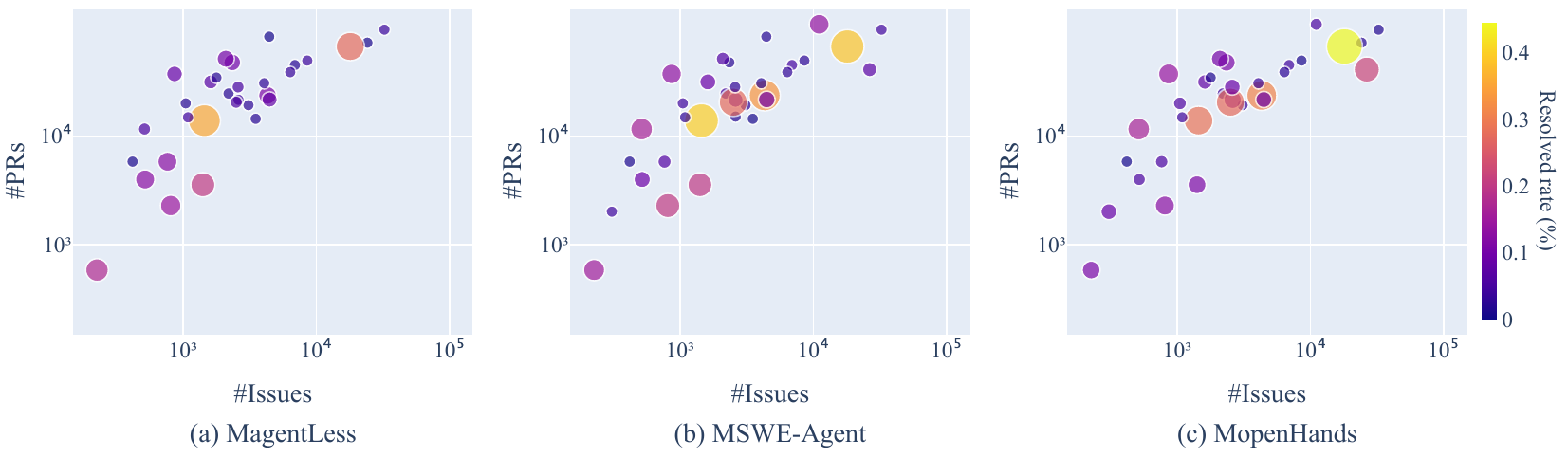}
    \vspace{-3mm}
    \caption{Relationship between resolved rate and the number of issues and PRs of a repository.}
    \label{fig:relation_with_issues_prs}
\end{figure}

\textbf{Performance across repositories of varying quality.}
To assess repository quality, we examine key metrics including the number of stars, forks, PRs, and issues. 
Fig.~\ref{fig:relation_with_repo_quality} illustrates the average resolved rate across LLMs for the three methods in relation to the number of stars and forks. 
Similarly, Fig.~\ref{fig:relation_with_issues_prs} shows the average resolved rate in relation to the number of issues and PRs. 
Both Fig.~\ref{fig:relation_with_repo_quality} and Fig.~\ref{fig:relation_with_issues_prs} exhibit a general positive correlation between \#Stars and \#Forks, as well as \#Issues and \#PRs across the majority of repositories.
Furthermore, repositories with higher resolved rates tend to cluster in the upper-right quadrant of both
Fig.~\ref{fig:relation_with_repo_quality} and Fig.~\ref{fig:relation_with_issues_prs},
suggesting that repositories with greater activity and community engagement (i.e., higher counts of stars, forks, issues, and PRs) are typically associated with a higher resolved rate. 
This trend is particularly evident for the \msweagent and \mopenhands.
In contrast, MagentLess exhibits relatively low variation in resolved rates across both Fig.~\ref{fig:relation_with_repo_quality} and Fig.~\ref{fig:relation_with_issues_prs},
underscoring an important observation: 
while a greater number of stars, forks, issues, and PRs tend to correlate with higher resolved rates, these metrics do not provide a guarantee of a repository’s issue-resolving effectiveness.

\begin{figure}[!ht]
    \centering
    \includegraphics[width=0.95\linewidth]{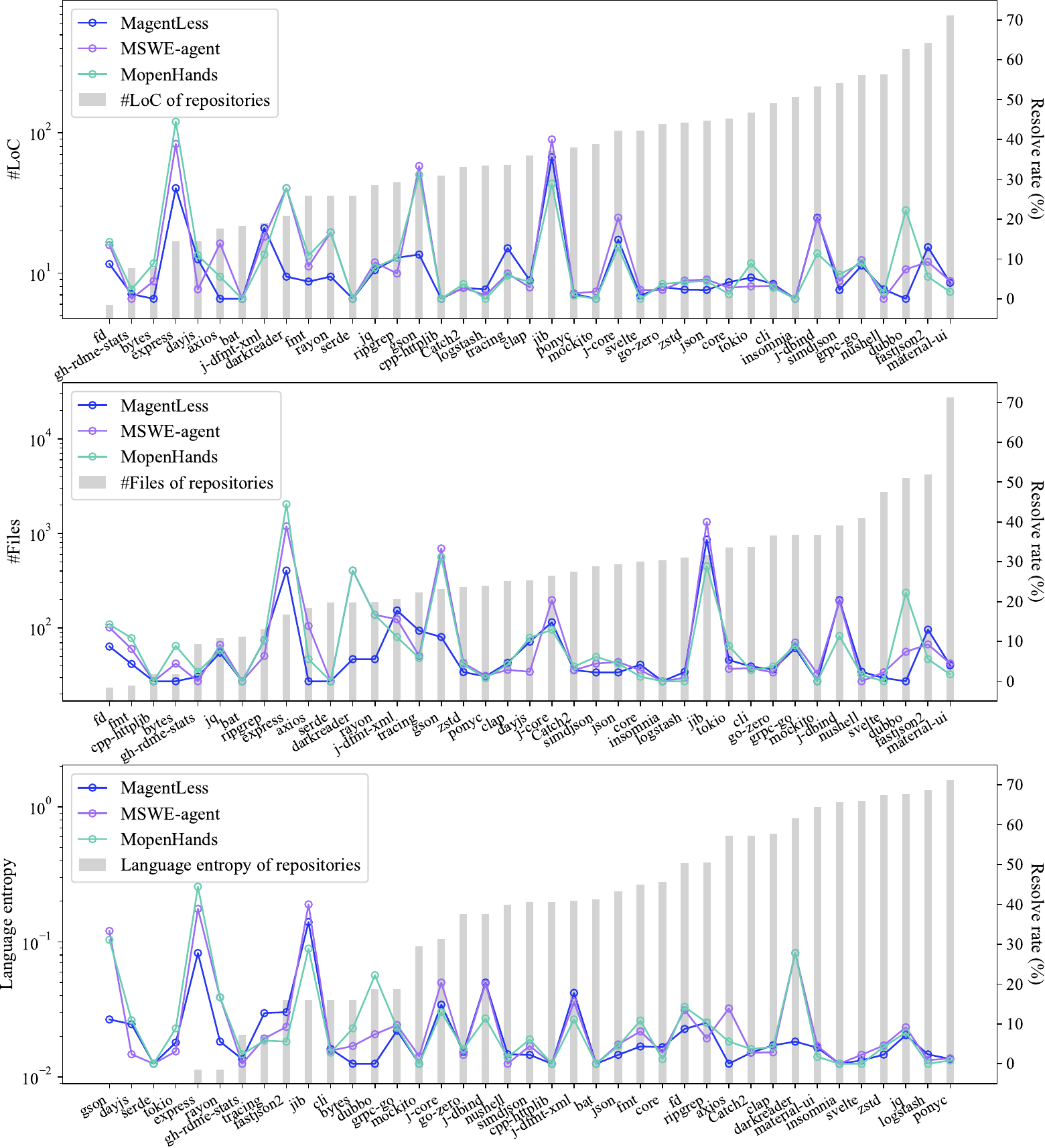}
    \caption{Relation between resolved rate and the repository complexity on \multiswebench.}
    \label{fig:relation_with_repo_complexity}
\end{figure}

\textbf{Performance across repositories with different levels of complexity.}
To evaluate repository complexity, we consider several key metrics: the number of lines of code (\#LoC), the number of files (\#Files), and language entropy. 
Let $L = \{l_1, l_2, \cdots,  l_n\}$ represent the set of programming languages used in the repository, with corresponding proportions $\{p_1, p_2, \cdots, p_n\}$. 
 The language entropy of the repository is then calculated as:
\begin{equation*}
    H(L) = - \sum_{i=1}^n p_i\log(p_i)
\end{equation*}
where $p_i$ denotes the proportion of the repository written in language $l_i$.
The average resolved rate across nine base LLMs with different repository complexity is presented in Fig.~\ref{fig:relation_with_repo_complexity}.

Fig.~\ref{fig:relation_with_repo_complexity} shows a consistent trend in the resolved rate across varied repository complexity: 
All three methods exhibit fluctuations in performance with changes in \#LoC, \#Files, and language entropy, generally decreasing as the repository complexity increases. 
For the impact of \#LoC, as \#LoC increases, the resolved rate tends to decrease. 
However, Java-based repositories, such as \href{https://github.com/google/gson}{\texttt{gson}}, \href{https://github.com/googlecontainertools/jib}{\texttt{jib}}, \href{https://github.com/fasterxml/jackson-core}{\texttt{j-core}}, \href{https://github.com/fasterxml/jackson-databind}{\texttt{j-dbind}}, and \href{https://github.com/apache/dubbo}{\texttt{dubbo}}, show higher resolved rates despite their larger size.
This suggests that factors beyond code size, such as lower language entropy, modularity, well-documented code, and adherence to standardized practices, play a significant role in improving performance.
For example, the \href{https://github.com/google/gson}{\texttt{gson}} repository demonstrates nearly-zero language entropy in Fig.~\ref{fig:relation_with_repo_complexity}.
Similarly, the impact of \#Files follows a trend similar to \#LoC. 
The impact of language entropy shows a clearer trend than that of \#LoC and \#Files: repositories with lower entropy typically achieve higher resolved rates. 
This indicates that code simplicity and consistency play a crucial role in improving issue-resolving effectiveness on a repository.

\subsection{Influencing Factors of Performance}
In this subsection, we investigate the factors influencing issue resolving performance, focusing on three key factors: 
(1) \emph{issue type}, examining how different types of issues impact resolving effectiveness;
(2) \emph{issue description characteristics}, evaluating the role of description length in resolving issues;
and (3) \emph{fix patch characteristics}, analyzing how the length of fix patches and the number of involved files influence the resolving performance.

\subsubsection{Issue Type}
\label{sec:issue_type}

Tab.~\ref{tab:issue_type} lists the performance of the three methods on \multiswebench across different issue types and languages.
Through a meticulous manual analysis of the annotation results in Sec.~\ref{sec:phase5_manual_verification}, we categorized all instances in \multiswebench into three issue types: bug fix (Bug Fix), new feature (New Feat.), and feature optimization (Feat. Opt.).
We observe a consistent performance hierarchy across all methods and languages: bug fix issues are resolved with the highest success rates, followed by new features, with feature optimization being the most challenging.
For instance, \msweagent achieves $17.97$\% on Java bug fixes but drops to $3.91$\% and $1.56$\% for new features and optimizations, respectively.
\magentless and \mopenhands show a similar trend in all languages.
These results highlight a fundamental limitation of current agent-based methods: they are more effective at localized, symptom-driven repairs, but struggle with semantically demanding tasks such as implementing new functionality or refining existing behavior.
The latter requires deeper intent understanding, multi-component reasoning, and cross-file context aggregation capabilities that remain underdeveloped in current LLM-based agents.

\begin{table*}[!ht]
    \centering
    \caption{Resolved rate(\%) on \multiswebench across different issue types (Claude-3.7-Sonnet).}
    \vspace{-2mm}
    \scalebox{0.7}{
    \begin{tabular}{l|ccc|ccc|ccc}
    \toprule
    \multirow{2}{*}{\textbf{Languages}} & \multicolumn{3}{c|}{\textbf{\magentless}} & \multicolumn{3}{c|}{\textbf{\msweagent}} & \multicolumn{3}{c}{\textbf{\mopenhands}} \\
    \cline{2-10}
       & Bug Fix & New Feat. & Feat. Opt. & Bug Fix & New Feat. & Feat. Opt. & Bug Fix & New Feat. & Feat. Opt.  \\
    \hline
    Java &  10.94 & 2.34 & 0.78 & 17.97 & 3.91 & 1.56 & 17.97 & 3.12 & 0.78  \\ 
    TypeScript &  2.68 & 0.45 & 0.45 & 9.38 & 1.34 & 0.45 & 1.79 & 0.00 & 0.45 \\ 
    JavaScript &  1.97 & 0.00 & 0.00 & 4.21 & 0.56 & 0.00 & 3.65 & 1.12 & 0.28\\ 
    Go &  3.74 & 0.93 & 1.17 & 3.27 & 0.70 & 1.40 &4.44 & 2.10 & 0.93 \\ 
    Rust &  4.60 & 0.42 & 0.42 & 5.44 & 1.26 & 0.00  & 12.97 & 2.93 & 0.00\\ 
    C &  6.25 & 0.00 & 0.00 & 7.81 & 0.78 & 0.00 & 7.81 & 0.78 & 0.00\\ 
    C++ & 2.33 & 0.78 & 0.00 & 7.75 & 3.1 & 0.78  &10.85 & 3.10 & 0.78 \\
    \bottomrule
    \end{tabular}}
    \label{tab:issue_type}
\end{table*}

\subsubsection{Characteristics of Issue Description}
\begin{figure}[ht]
    \centering
    \includegraphics[width=0.7\linewidth]{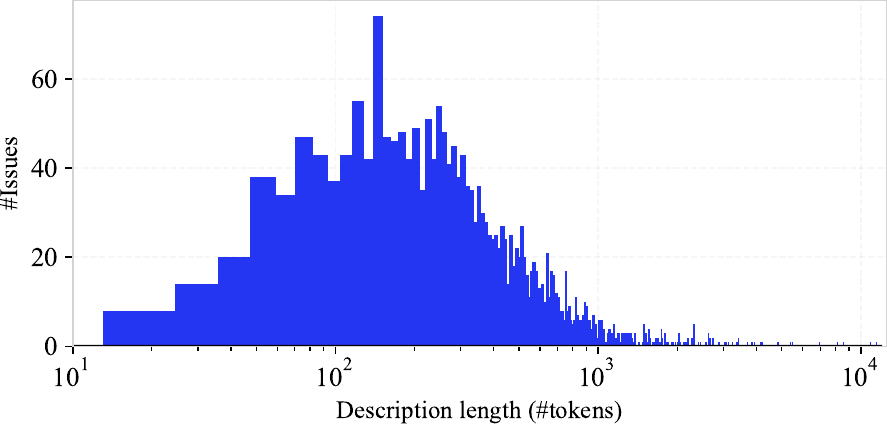}
    \caption{Histogram of issue description length (\#tokens).}
\label{fig:histogram}
\end{figure}

\begin{figure}[t]
    \centering
    \includegraphics[width=1\linewidth]{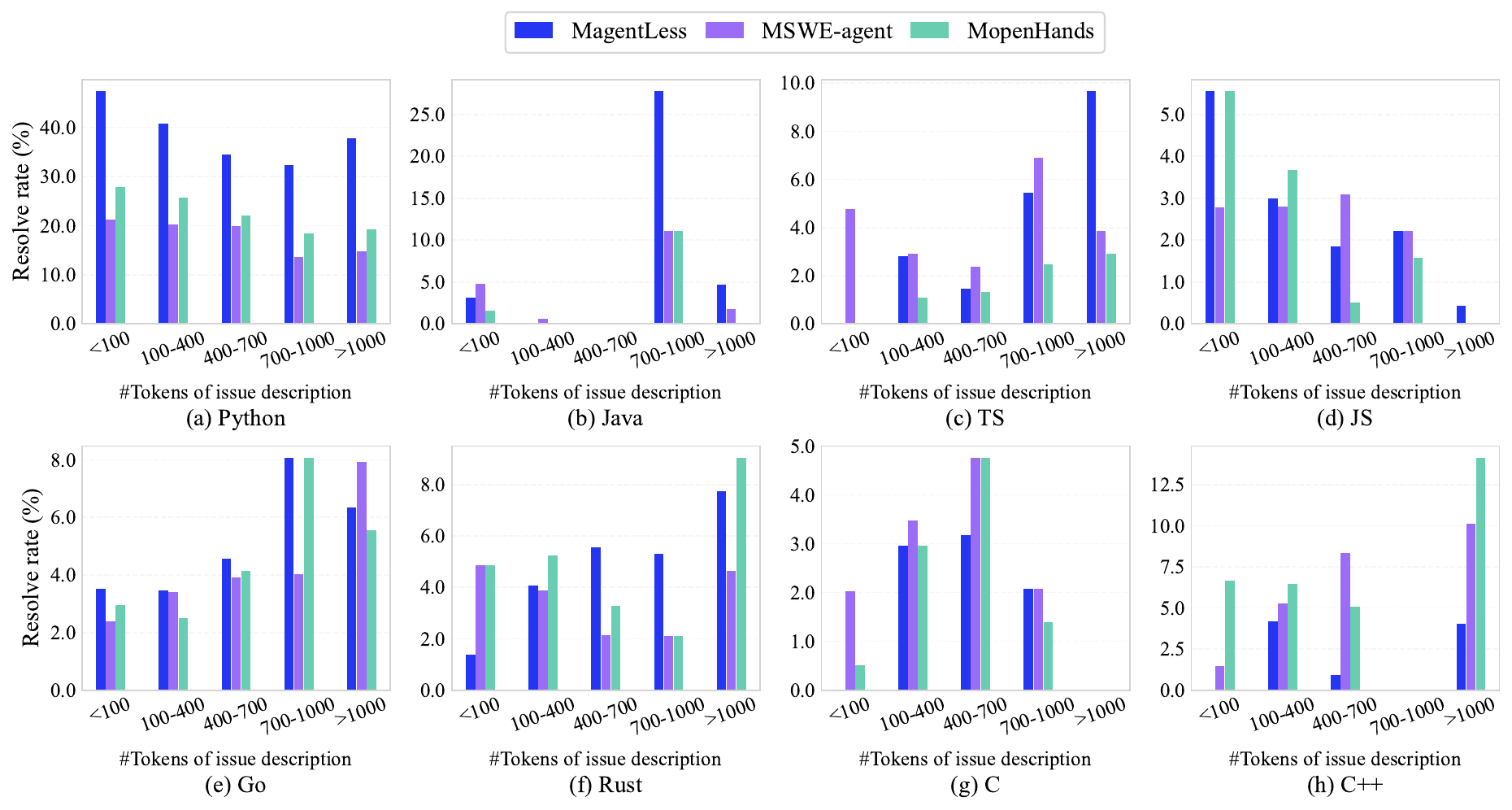}
    \caption{Influence of issue description length on resolved rate.}
\label{fig:issue_length}
\end{figure}

We aim to examine the impact of issue description length on issue-resolving performance. 
Fig.~\ref{fig:histogram} illustrates the distribution of issue lengths (in tokens) in \multiswebench, which follows a power law, with the majority of issues being under 1,000 tokens. 
To explore the effect of description length, the issues are categorized into 5 intervals: <100, 100-400, 400-700, 700-1000, and >1000 tokens, as shown in Fig.~\ref{fig:issue_length}.
The absence of a corresponding bars indicates cases where no issues are successfully resolved.

As shown in Fig.~\ref{fig:issue_length}, there is no consistent relationship between issue description length and resolved rate.
For example, in Python, issues with longer descriptions tend to have lower resolved rates, whereas in Go, longer descriptions are associated with higher rates.
This discrepancy arises from two potential types of long issue descriptions: (1) detailed issues with precise issue position indications and resolving steps, and (2) complex issues that require extended descriptions to explain. 
These two possibilities have distinct impacts on the difficulty of resolving an issue, influencing the resolved rate in different ways.
As for the performance among methods, compared with MSWE-agent and MopenHands, MagentLess generally performs better with longer, more detailed descriptions on average of the nine LLMs, especially for languages like Python, Java, and TS.

\subsubsection{Characteristics of Fix Patches}

\begin{figure}[th]
    \centering
    \begin{minipage}[b]{0.48\linewidth}
        \centering
        \includegraphics[width=1\linewidth]{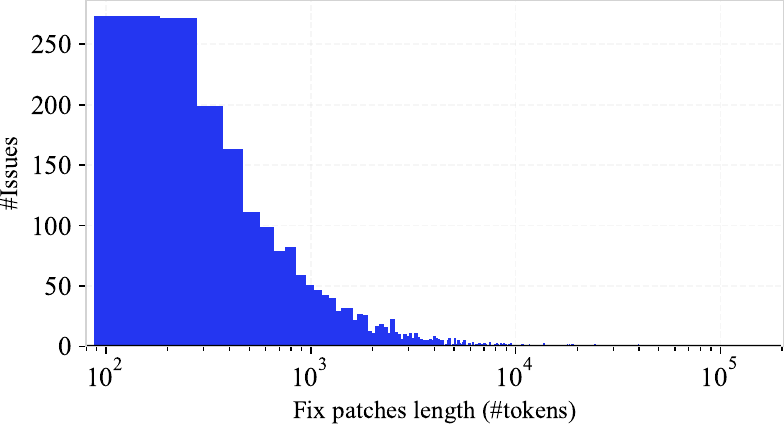}
        \caption{Histogram of fix patches length (\#tokens).}
        \label{fig:histogram_fix_patch_length}
    \end{minipage}
    \hfill
    \begin{minipage}[b]{0.48\linewidth}
        \centering
        \includegraphics[width=1\linewidth]{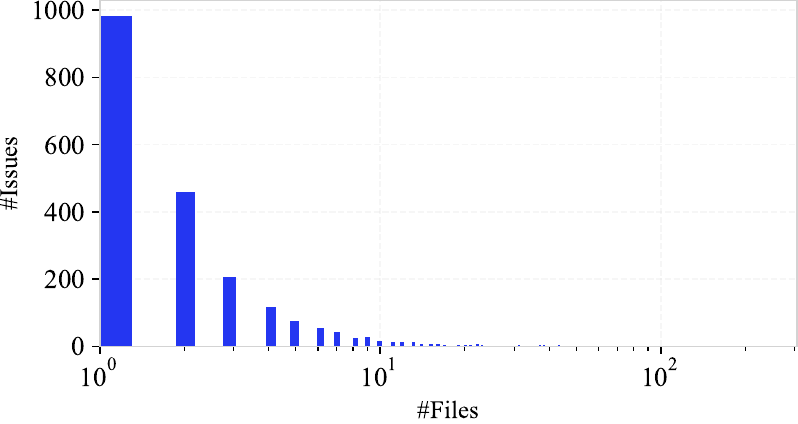}
        \caption{Histogram of the number of files modified by fix patches.}
        \label{fig:histogram_num_files}
    \end{minipage}
\end{figure}

In this subsection, we investigate the impact of the ground-truth fix patches on the resolved rate, focusing on two key factors: 
(1) \emph{Fix patch length}:
We analyze how the length of fix patches affects performance, noticing that longer patches require more complex reasoning capabilities from LLMs. 
The fix patches are categorized into five intervals based on the length distribution shown in Fig.~\ref{fig:histogram_fix_patch_length}: <200, 200-600, 600-1000, 1000-1400, and >1400 tokens.
(2) \emph{Number of files modified by fix patches}: We examine how the cross-file nature of the fix patches influences performance, with more files requiring enhanced cross-file handling capabilities. The number of modified files is divided into four categories: 1, 1-5, 5-10, and >10, with the distribution shown in Fig.\ref{fig:histogram_num_files}.
The absence of corresponding bars indicates cases where no issues are successfully resolved.

\begin{figure}[t]
    \centering
    \includegraphics[width=1\linewidth]{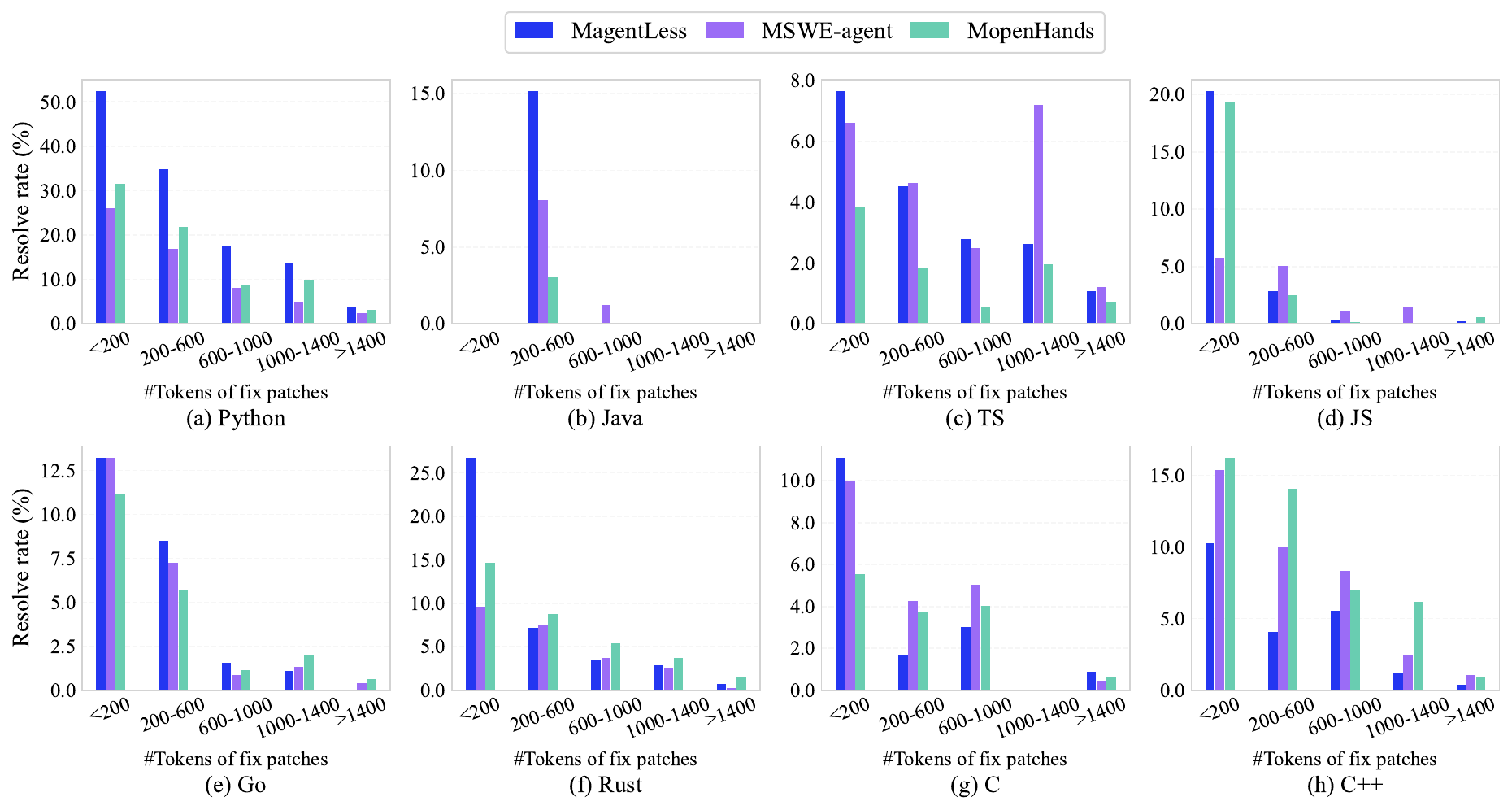}
    \caption{Influence of fix patch length on resolved rate.}
    \label{fig:relation_with_fix_patch_length}
\end{figure}

\textbf{Performance drops as fix patch length increases.}
As shown in Fig.~\ref{fig:relation_with_fix_patch_length}, the length of fix patches significantly impacts the resolved rate, with shorter patches generally leading to higher success rates.
Specifically, in the majority of cases, issues with descriptions >600 tokens exhibit a resolved rate approximately 50\% lower than that of issues with descriptions <200 tokens.
For most programming languages, the resolved rate for shorter fix patches is notably higher, especially for \magentless, which shows a peak in this range for languages like Python, Java, and C. 
This suggests that shorter patches are easier to handle, as they require less reasoning and simpler edits. 
For all three methods, the resolved rate for very long fix patches (>1000 tokens) drops significantly, even reaching zero for Java. This indicates that long patches, which likely require handling a larger scope of modifications, present greater challenges, especially for methods that may not be optimized for such complex tasks.

\begin{figure}[t]
    \centering
    \includegraphics[width=1\linewidth]{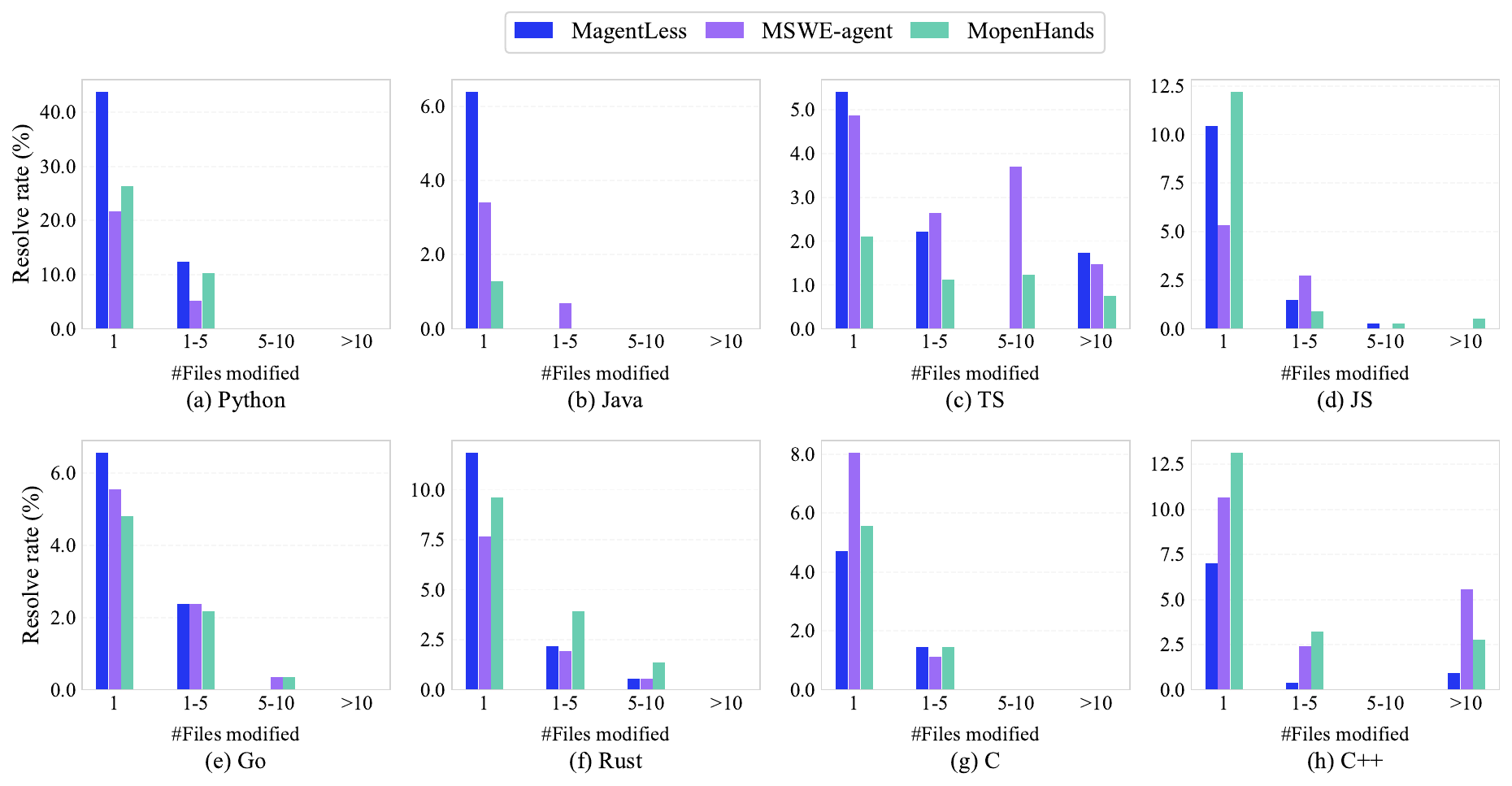}
    \caption{Influence of the number of files modified by fix patches on resolved rate.}
    \label{fig:relation_with_fix_patch_files}
\end{figure}

\textbf{Cross-file fix patches lead to reduced effectiveness.}
Fig.~\ref{fig:relation_with_fix_patch_files} illustrates the relationship between the number of files modified by fix patches and the resolved rate. 
Consistent with the observation in Fig.~\ref{fig:relation_with_fix_patch_length}, resolved rate drops significantly as the number of modified files increases across all three methods.
This trend highlights the potential challenge of understanding and resolving issues that require changes across multiple files, which may demand more intricate handling or coordination between different parts of the repository.
For issues resolved by modifications in a single file, \magentless outperforms \msweagent and \mopenhands in five out of seven programming languages. This suggests that \magentless is more effective at resolving issues within the scope of a single file.


\subsection{Case Study}
\label{sec:case_study}
In this subsection, we analyze representative cases that highlight the strengths of agents, common failure patterns, and language-specific challenges, providing insights for future directions.

\noindent\textit{\textbf{6.3.1 Language-General Case}}
\vspace{-0.3cm}
\begin{itemize}[leftmargin=*]
    \item \msweagent and \mopenhands often failed by exhausting the $50$-round interaction limit, sometimes without even triggering the submit action, as seen in cases like \href{https://github.com/multi-swe-bench/experiments/blob/main/evaluation/javascript/verified/20250329_MSWE-agent_DeepSeek-V3/trajs/axios\_\_axios-5919.traj}{axios\_\_axios-5919.traj}, \href{https://github.com/multi-swe-bench/experiments/blob/main/evaluation/rust/verified/20250329_MSWE-agent_DeepSeek-V3/trajs/clap-rs\_\_clap-5520.traj}{clap-rs\_\_clap-5520.traj}, and \href{https://github.com/multi-swe-bench/experiments/blob/main/evaluation/go/verified/20250329_MopenHands_DeepSeek-R1/trajs/cli\_\_cli-513/DeepSeek-R1-1743467902.7243466.json}{cli\_\_cli-513.traj}.
    Future work may explore strategies that enable agents to solve more complex tasks within a limited number of interaction rounds.
    \item A significant number of failures across all three agent methods were due to incorrect fault localization, which led to an inability to identify and modify the relevant code, as seen in cases such as \href{https://github.com/multi-swe-bench/experiments/blob/main/evaluation/java/verified/20250329_MSWE-agent_DeepSeek-V3/trajs/elastic\_\_logstash-14898.traj}{elastic\_\_logstash-14898.traj},  \href{https://github.com/multi-swe-bench/experiments/blob/main/evaluation/java/verified/20250329_MSWE-agent_DeepSeek-V3/trajs/alibaba\_\_fastjson2-2285.traj}{alibaba\_\_fastjson2-2285.traj}, \href{https://github.com/multi-swe-bench/experiments/blob/main/evaluation/java/verified/20250329_MagentLess_DeepSeek-V3/trajs/fasterxml\_\_jackson-databind-3560.traj}{fasterxml\_\_jackson-databind-3560.traj}, and  \href{https://github.com/multi-swe-bench/experiments/blob/main/evaluation/java/verified/20250329_MopenHands_Claude-3.7-Sonnet/trajs/apache\_\_dubbo-7041/Claude-3.7-Sonnet-1743283670.1144483.json}{apache\_\_dubbo-7041.traj}.
    This highlights the centrality of accurate fault localization and points to the potential of integrating software engineering techniques like SBFL~\citep{sbfl_1,sbfl_2} into future agent designs.
    \item In cases such as \href{https://github.com/multi-swe-bench/experiments/blob/main/evaluation/python/verified/20250329_SWE-agent_DeepSeek-V3/trajs/astropy__astropy-12907.traj}{astropy\_\_astropy-12907.traj} and \href{https://github.com/multi-swe-bench/experiments/blob/main/evaluation/python/verified/20250329_SWE-agent_DeepSeek-V3/trajs/django__django-11299.traj}{django\_\_django-11299.traj}, the model generated multiple valid actions in a single turn, but the hardcoded agent framework executed only the last, resulting in premature submission. This reveals a structural bottleneck in current agent design, where rigid control logic overrides model intent. It calls for a shift toward lightweight, model-centric agents with full decision autonomy delegated to the LLM.
    \item Bug reproduction plays a critical role in successful repair. In cases such as \href{https://github.com/multi-swe-bench/experiments/blob/main/evaluation/c++/verified/20250329_MSWE-agent_Claude-3.5-Sonnet(Oct)/trajs/nlohmann\_\_json-4537.traj}{nlohmann\_\_json-4537.traj},  \href{https://github.com/multi-swe-bench/experiments/blob/main/evaluation/c++/verified/20250329_MSWE-agent_Claude-3.5-Sonnet(Oct)/trajs/fmtlib\_\_fmt-3248.traj}{fmtlib\_\_fmt-3248.traj}, \href{https://github.com/multi-swe-bench/experiments/blob/main/evaluation/java/verified/20250329_MopenHands_Claude-3.5-Sonnet(Oct)/trajs/fasterxml\_\_jackson-core-1142/Claude-3.5-Sonnet(Oct)-1740475152.28328.json}{fasterxml\_\_jackson-core-1142.traj}, and \href{https://github.com/multi-swe-bench/experiments/blob/main/evaluation/java/verified/20250329_MopenHands_Claude-3.5-Sonnet(Oct)/trajs/google\_\_gson-1093/Claude-3.5-Sonnet(Oct)-1740478999.7829874.json}{google\_\_gson-1093.traj}, the model successfully reproduced the issue before producing an effective fix. In contrast, failure to reproduce often resulted in unresolved cases, as seen in \href{https://github.com/multi-swe-bench/experiments/blob/main/evaluation/c++/verified/20250329_MSWE-agent_Claude-3.5-Sonnet(Oct)/trajs/catchorg\_\_Catch2-1609.traj}{catchorg\_\_Catch2-1609.traj}.
    However, reproduction is not always a prerequisite for success.
    Claude-3.5-Sonnet and Claude-3.7-Sonnet occasionally bypass reproduction and edit the code directly—yet still resolve the issue successfully, as in \href{https://github.com/multi-swe-bench/experiments/blob/main/evaluation/c++/verified/20250329_MSWE-agent_Claude-3.5-Sonnet(Oct)/trajs/nlohmann\_\_json-3601.traj}{nlohmann\_\_json-3601.traj}, \href{https://github.com/multi-swe-bench/experiments/blob/main/evaluation/c++/verified/20250329_MSWE-agent_Claude-3.5-Sonnet(Oct)/trajs/fmtlib\_\_fmt-3729.traj}{fmtlib\_\_fmt-3729.traj}, and \href{https://github.com/multi-swe-bench/experiments/blob/main/evaluation/java/verified/20250329_MopenHands_Claude-3.5-Sonnet(Oct)/trajs/googlecontainertools\_\_jib-4035/Claude-3.5-Sonnet(Oct)-1740482730.20658.json}{googlecontainertools\_\_jib-4035.traj}. 
    These cases suggest that agents should selectively invoke reproduction based on factors such as error traceability, edit confidence, and execution cost.
\end{itemize}

\noindent\textit{\textbf{6.3.2 Language-Specific Case}}
\vspace{-0.3cm}
\begin{itemize}[leftmargin=*]
    \item For certain TypeScript projects, the length of the extracted repository structure often exceeds the model's maximum context length, preventing \magentless from performing fault localization
    (e.g., \href{https://github.com/multi-swe-bench/experiments/blob/main/evaluation/typescript/verified/20250329_MagentLess_Doubao-1.5-pro/trajs/mui\_\_material-ui-25852.traj}{mui\_\_material-ui-25852.traj} and \href{https://github.com/multi-swe-bench/experiments/blob/main/evaluation/typescript/verified/20250329_MagentLess_Doubao-1.5-pro/trajs/mui\_\_material-ui-37850.traj}{mui\_\_material-ui-37850.traj}).
    This reveals the limited generalizability of fixed workflows like \magentless when confronted with structurally irregular and language-specific scenarios, indicating significant room for improvement in both robustness and adaptability.
    \item Tree-sitter fails to reliably extract code structures in JavaScript repositories that use loosely bound syntax such as arrow functions, preventing \magentless from constructing contextual windows around candidate edits (e.g., \href{https://github.com/multi-swe-bench/experiments/blob/main/evaluation/javascript/verified/20250329_MagentLess_Claude-3.7-Sonnet/trajs/iamkun\_\_dayjs-2532.traj}{iamkun\_\_dayjs-2532.traj} and \href{https://github.com/multi-swe-bench/experiments/blob/main/evaluation/javascript/verified/20250329_MagentLess_Claude-3.7-Sonnet/trajs/iamkun\_\_dayjs-2399.traj}{iamkun\_\_dayjs-2399.traj}).
    This exposes a structural brittleness in syntax-driven workflows when applied to syntactically permissive languages, motivating future extensions of \magentless toward greater tolerance to parsing failure and language-specific irregularities.
    \item In some JavaScript projects, agents sometimes invoke \texttt{pnpm} to launch development servers as part of the repair routine.
    However, current agent frameworks lack support for managing long-lived, interactive processes, often resulting in premature termination or container crashes (e.g., \href{https://github.com/multi-swe-bench/experiments/blob/main/evaluation/javascript/verified/20250329_MSWE-agent_OpenAI-o1/trajs/sveltejs\_\_svelte-12460.traj}{sveltejs\_\_svelte-12460.traj} and \href{https://github.com/multi-swe-bench/experiments/blob/main/evaluation/javascript/verified/20250329_MSWE-agent_OpenAI-o1/trajs/sveltejs\_\_svelte-10077.traj}{sveltejs\_\_svelte-10077.traj)}.
    Future agents should support persistent shell sessions and interactive service control, as enabled by frameworks like SWE-ReX~\citep{swe_rex}.
\end{itemize}

\subsection{Resource Consumption}

\begin{table*}[t]
    \centering
    \vspace{-3mm}
    \caption{Average token consumption on \multiswebench. In. represents the average number of input tokens (in thousands), and Out. is the average number of output tokens (in thousands).}
    \label{tab:consumption}
    \scalebox{0.61}{
    \begin{tabular}{l|cc|cc|cc|cc|cc|cc|cc|cc} 
    \toprule
    \multirow{2}{*}{\textbf{Models}} & \multicolumn{2}{c|}{\textbf{Python}} & \multicolumn{2}{c|}{\textbf{Java}} & \multicolumn{2}{c|}{\textbf{TS}} & \multicolumn{2}{c|}{\textbf{JS}} & \multicolumn{2}{c|}{\textbf{Go}} & \multicolumn{2}{c|}{\textbf{Rust}} & \multicolumn{2}{c|}{\textbf{C}} & \multicolumn{2}{c}{\textbf{C++}} \\
    \cmidrule{2-17}
    & In. & Out. & In. & Out. & In. & Out. & In. & Out. & In. & Out. & In. & Out. & In. & Out. & In. & Out. \\
    \hline
    \rowcolor{mygray}\multicolumn{17}{c}{\textbf{\magentless}} \\
    GPT‑4o & 36.15 & 4.20 & 52.10 & 2.74 & 241.18 & 2.01 & 29.48 & 2.53 & 25.14 & 2.72 & 48.23 & 2.71 & 50.26 & 2.64 & 76.38 & 2.44\\
    OpenAI-o1 & 34.43 & 3.76 & 50.18 & 1.92 & 240.47 & 1.21 & 36.53 & 1.26 & 24.51 & 1.70 & 58.59 & 1.49 & 48.08 & 1.57 & 119.64 & 1.31\\
    OpenAI-o3-mini-high & 31.38 & 4.50 & 79.48 & 2.36 & 245.39 & 1.54 & 38.28 & 1.55 & 31.58 & 1.67 & 68.80 & 2.05 & 73.48 & 1.99 & 200.29 & 1.91\\
    Claude-3.5-Sonnet & 39.13 & 5.38 & 48.42 & 2.67 & 239.93 & 1.86 & 28.46 & 2.39 & 22.80 & 2.79 & 51.25 & 2.66 & 49.13 & 2.55 & 96.85 & 2.49\\
    Claude-3.7-Sonnet & 27.99 & 6.54 & 63.97 & 3.16 & 248.36 & 2.08 & 26.66 & 3.17 & 22.79 & 3.63 & 81.15 & 3.33 & 50.52 & 3.19 & 129.34 & 2.91\\
    DeepSeek-V3 & 39.97 & 4.26 & 42.35 & 2.70 & 244.32 & 1.92 & 26.44 & 2.51 & 22.78 & 2.65 & 83.04 & 2.47 & 92.53 & 2.38 & 189.08 & 2.11\\
    DeepSeek-R1 & 31.35 & 2.80 & 70.35 & 1.76 & 249.02 & 1.10 & 28.23 & 1.30 & 21.69 & 1.79 & 88.73 & 1.52 & 100.99 & 1.39 & 177.66 & 1.41\\
    Qwen2.5-72B-Instruct & 28.60 & 3.46 & 62.95 & 2.52 & 243.98 & 1.65 & 26.11 & 2.14 & 24.67 & 3.36 & 50.63 & 2.89 & 55.93 & 2.78 & 150.44 & 2.19\\
    Doubao-1.5-pro & 42.75 & 2.91 & 116.09 & 1.36 & 249.51 & 1.55 & 36.37 & 2.07 & 29.38 & 3.15 & 124.67 & 1.62 & 121.94 & 1.52 & 216.21 & 0.76\\
    \midrule
    \rowcolor{mygray}\multicolumn{17}{c}{\textbf{\msweagent}} \\
    GPT‑4o & 166.91 & 3.08 & 51.05 & 4.54 & 46.39 & 4.63 & 32.01 & 4.36 & 36.73 & 4.71 & 43.79 & 4.71 & 39.47 & 4.57 & 55.49 & 4.96\\
    OpenAI-o1 & 243.44 & 1.64 & 33.36 & 2.99 & 30.05 & 3.56 & 25.70 & 3.19 & 37.71 & 3.49 & 39.51 & 3.71 & 34.05 & 3.17 & 29.24 & 3.28\\
    OpenAI-o3-mini-high & 240.23 & 1.82 & 26.37 & 3.64 & 18.27 & 3.39 & 21.33 & 3.99 & 26.46 & 3.41 & 32.84 & 4.03 & 23.24 & 3.78 & 32.39 & 4.90\\
    Claude-3.5-Sonnet & 33.30 & 5.55 & 32.09 & 3.89 & 21.51 & 3.10 & 23.94 & 3.66 & 21.06 & 3.06 & 35.47 & 4.03 & 31.16 & 3.44 & 38.22 & 4.32\\
    Claude-3.7-Sonnet & 31.86 & 4.46 & 38.96 & 4.79 & 32.08 & 4.92 & 32.16 & 4.60 & 33.79 & 4.56 & 40.59 & 4.56 & 38.41 & 4.34 & 36.96 & 4.67\\
    DeepSeek-V3 & 12.63 & 22.83 & 35.08 & 4.14 & 15.73 & 2.15 & 19.78 & 3.23 & 15.34 & 2.43 & 33.98 & 5.47 & 16.26 & 2.07 & 31.28 & 4.18\\
    DeepSeek-R1 & 11.76 & 2.65 & 17.51 & 2.69 & 9.91 & 1.66 & 9.36 & 2.43 & 10.47 & 1.85 & 13.98 & 2.86 & 11.34 & 2.44 & 14.64 & 3.06\\
    Qwen2.5-72B-Instruct & 164.42 & 6.69 & 53.43 & 9.26 & 39.58 & 7.82 & 35.21 & 6.45 & 22.53 & 8.38 & 36.49 & 7.93 & 28.90 & 5.76 & 67.29 & 11.69\\
    Doubao-1.5-pro & 72.58 & 1.30 & 37.75 & 3.73 & 19.18 & 2.46 & 32.90 & 3.68 & 25.39 & 3.91 & 38.09 & 4.04 & 29.03 & 3.65 & 32.67 & 3.59\\
    \midrule
    \rowcolor{mygray}\multicolumn{17}{c}{\textbf{\mopenhands}} \\
    GPT‑4o & 25.35 & 1.24 & 22.01 & 1.32 & 35.76 & 1.60 & 35.51 & 1.50 & 23.96 & 1.52 & 40.40 & 1.45 & 34.80 & 2.08 & 34.61 & 1.66\\
    OpenAI-o1 & 19.27 & 1.20 & 18.69 & 1.27 & 27.28 & 2.14 & 30.96 & 2.07 & 21.09 & 1.56 & 28.90 & 1.55 & 27.18 & 1.67 & 21.55 & 1.57\\
    OpenAI-o3-mini-high & 21.52 & 5.18 & 22.82 & 3.88 & 30.70 & 4.32 & 36.57 & 4.06 & 25.44 & 4.14 & 30.64 & 3.15 & 23.98 & 3.26 & 23.76 & 4.26\\
    Claude-3.5-Sonnet & 32.35 & 7.69 & 31.97 & 5.35 & 35.88 & 6.43 & 38.91 & 6.14 & 27.31 & 7.26 & 55.51 & 6.23 & 55.79 & 5.66 & 35.85 & 6.74\\
    Claude-3.7-Sonnet & 26.04 & 7.84 & 28.43 & 7.86 & 31.06 & 7.31 & 38.06 & 7.46 & 30.05 & 7.86 & 48.30 & 7.00 & 35.25 & 6.30 & 33.14 & 7.76\\
    DeepSeek-V3 & 18.97 & 5.16 & 26.35 & 3.65 & 26.60 & 3.69 & 29.08 & 4.43 & 15.42 & 3.31 & 32.90 & 5.05 & 21.77 & 3.53 & 30.67 & 5.36\\
    DeepSeek-R1 & 11.25 & 5.13 & 17.15 & 5.04 & 12.71 & 4.33 & 17.85 & 5.29 & 12.65 & 5.14 & 17.58 & 6.95 & 24.16 & 6.11 & 17.38 & 7.62\\
    Qwen2.5-72B-Instruct & 27.28 & 10.38 & 33.26 & 11.80 & 36.86 & 9.12 & 28.84 & 9.07 & 21.17 & 11.35 & 37.14 & 10.99 & 35.02 & 9.69 & 35.34 & 10.88\\
    Doubao-1.5-pro & 23.16 & 3.95 & 24.15 & 3.35 & 18.34 & 1.66 & 23.75 & 2.76 & 18.21 & 3.78 & 27.40 & 2.07 & 26.54 & 2.82 & 26.07 & 3.44\\
    \bottomrule
    \end{tabular}
}
\end{table*}

In this subsection, we analyze the resource consumption across different languages, focusing on two key metrics: (1) the average token consumption and (2) the average cost per issue.

\textbf{Average token consumption per issue.}
Tab.~\ref{tab:consumption} compares the average token consumption for various languages using the GPT-4o tokenizer.
Overall, token consumption varies between methods and languages. 
Among languages, TS exhibits the highest token consumption in \magentless, whereas Python is the most token-intensive language in \msweagent. 
Notably, Go demonstrates relatively low token consumption in both input and output, likely due to its minimalistic syntax and clear conventions, which contribute to its compact representation and reduced token overhead. 
Additionally, in \msweagent for Python, we observe increased token usage on LLMs, including GPT-4o, OpenAI-o1, OpenAI-o3-mini-high, and Qwen2.5-72B-Instruct. 
This is because we maintain the original \sweagent implementation for Python, which does not incorporate the over-length truncation mechanism applied to other languages.

\begin{table*}[th]
    \centering
    \renewcommand{\arraystretch}{1.1}
    \caption{Average cost (\$) per issue of different models and methods on \multiswebench.}
    \vspace{-3mm}
    \label{tab:cost}
    \scalebox{0.75}{
    \begin{tabular}{lcccccccccccccc} 
    \toprule
    \textbf{Models} & \textbf{Python} & \textbf{Java} & \textbf{TS} & \textbf{JS} & \textbf{Go} & \textbf{Rust} & \textbf{C} & \textbf{C++} \\
    \hline
    \rowcolor{mygray}\multicolumn{9}{c}{\textbf{\magentless}} \\
    GPT-4o & 0.1324 & 0.1576 & 0.6230 & 0.0990 & 0.0900 & 0.1476 & 0.1520 & 0.2153\\
    OpenAI-o1 & 0.7417 & 0.8680 & 3.6795 & 0.6233 & 0.4698 & 0.9682 & 0.8153 & 1.8734\\
    OpenAI-o3-mini-high & 0.0543 & 0.0978 & 0.2767 & 0.0489 & 0.0421 & 0.0847 & 0.0896 & 0.2287\\
    Claude-3.5-Sonnet & 0.1981 & 0.1853 & 0.7478 & 0.1213 & 0.1102 & 0.1937 & 0.1856 & 0.3280\\
    Claude-3.7-Sonnet & 0.1821 & 0.2393 & 0.7763 & 0.1275 & 0.1229 & 0.2933 & 0.1994 & 0.4317\\
    DeepSeek-V3 & 0.0075 & 0.0059 & 0.0192 & 0.0046 & 0.0045 & 0.0085 & 0.0091 & 0.0156\\
    DeepSeek-R1 & 0.0105 & 0.0137 & 0.0373 & 0.0068 & 0.0070 & 0.0158 & 0.0172 & 0.0280\\
    Qwen2.5-72B-Instruct & 0.0051 & 0.0092 & 0.0324 & 0.0042 & 0.0046 & 0.0077 & 0.0084 & 0.0204\\
    Doubao-1.5-pro & 0.0055 & 0.0132 & 0.0279 & 0.0046 & 0.0041 & 0.0142 & 0.0138 & 0.0240\\
    \midrule
    \rowcolor{mygray}\multicolumn{9}{c}{\textbf{\msweagent}} \\
    GPT-4o & 0.4480 & 0.1731 & 0.1623 & 0.1236 & 0.1390 & 0.1565 & 0.1444 & 0.1883\\
    OpenAI-o1 & 3.7499 & 0.6797 & 0.6644 & 0.5772 & 0.7749 & 0.8151 & 0.7010 & 0.6353\\
    OpenAI-o3-mini-high & 0.2722 & 0.0450 & 0.0350 & 0.0410 & 0.0441 & 0.0538 & 0.0422 & 0.0572\\
    Claude-3.5-Sonnet & 0.1831 & 0.1546 & 0.1110 & 0.1266 & 0.1091 & 0.1669 & 0.1451 & 0.1794\\
    Claude-3.7-Sonnet & 0.1626 & 0.1887 & 0.1700 & 0.1654 & 0.1698 & 0.1901 & 0.1803 & 0.1810\\
    DeepSeek-V3 & 0.0260 & 0.0070 & 0.0035 & 0.0049 & 0.0037 & 0.0084 & 0.0034 & 0.0068\\
    DeepSeek-R1 & 0.0075 & 0.0083 & 0.0050 & 0.0066 & 0.0055 & 0.0082 & 0.0069 & 0.0088\\
    Qwen2.5-72B-Instruct & 0.0241 & 0.0106 & 0.0083 & 0.0072 & 0.0063 & 0.0079 & 0.0061 & 0.0134\\
    Doubao-1.5-pro & 0.0083 & 0.0052 & 0.0028 & 0.0046 & 0.0039 & 0.0053 & 0.0042 & 0.0046\\
    \midrule
    \rowcolor{mygray}\multicolumn{9}{c}{\textbf{\mopenhands}} \\
    GPT-4o & 0.0758 & 0.0682 & 0.1054 & 0.1038 & 0.0751 & 0.1155 & 0.1078 & 0.1031\\
    OpenAI-o1 & 0.3608 & 0.3564 & 0.5374 & 0.5885 & 0.4099 & 0.5262 & 0.5081 & 0.4171\\
    OpenAI-o3-mini-high & 0.0465 & 0.0422 & 0.0528 & 0.0581 & 0.0462 & 0.0476 & 0.0407 & 0.0449\\
    Claude-3.5-Sonnet & 0.2124 & 0.1761 & 0.2041 & 0.2089 & 0.1908 & 0.2601 & 0.2523 & 0.2086\\
    Claude-3.7-Sonnet & 0.1957 & 0.2032 & 0.2028 & 0.2261 & 0.2080 & 0.2500 & 0.2002 & 0.2158\\
    DeepSeek-V3 & 0.0070 & 0.0059 & 0.0059 & 0.0069 & 0.0047 & 0.0079 & 0.0054 & 0.0080\\
    DeepSeek-R1 & 0.0128 & 0.0134 & 0.0113 & 0.0141 & 0.0130 & 0.0177 & 0.0168 & 0.0191\\
    Qwen2.5-72B-Instruct & 0.0077 & 0.0090 & 0.0084 & 0.0074 & 0.0073 & 0.0092 & 0.0084 & 0.0089\\
    Doubao-1.5-pro & 0.0037 & 0.0036 & 0.0025 & 0.0034 & 0.0031 & 0.0036 & 0.0037 & 0.0038\\
    \bottomrule
    \end{tabular}
}
\end{table*}

\textbf{Average cost per issue.}
Tab.~\ref{tab:cost} presents the average cost (\$) per issue on \multiswebench.
Notably, DeepSeek-V3, DeepSeek-R1, and Qwen2.5-72B-Instruct achieve the lowest cost per resolved issue, staying below \$0.03, benefiting from their cost-efficient pricing (below \$0.14 per million input tokens).
In contrast, OpenAI-o1 is the most expensive model, due to its high token price (\$15 per million input tokens).
Overall, \magentless exhibits lower costs compared to \msweagent by virtue of its fixed workflow.
Conversely, the workflows of both \msweagent and \mopenhands are determined by LLMs, requiring more interaction turns with the environment, which results in higher costs.

\subsection{Troubleshooting}
\label{sec:troubleshooting}
During the construction of \multiswebench and \multiswerl, we encountered a range of practical and non-obvious challenges.
We document the key issues below to facilitate reproducibility and guide future community contributions:
\begin{itemize}[leftmargin=*]
    \item Test log inconsistency. The number of test cases differs between Test.log and Fix.log, as fix.patch may optimize control flow, eliminate redundant coverage, or merge test paths, which is commonly observed in repositories such as \href{https://github.com/preactjs/preact}{preactjs/preact}.
    \item Pre-fix build failures. Certain repositories fail to compile or execute tests before applying fix.patch, due to newly introduced symbols (e.g., functions or variables) in test.patch that are undefined without the fix.
    \item Binary artifacts in C\&C++. Agent runs may generate compiled binaries (e.g., "\texttt{.o}", "\texttt{.bin}") that block "\texttt{git apply}". We currently strip these via hard-coded filtering, though more robust handling is needed.
    \item Evaluation nondeterminism. Java and C tests occasionally exhibit unstable behavior due to excessive thread concurrency, leading to inconsistent run.log outcomes. We mitigate this by reducing parallelism during evaluation.
    \item Name casing mismatches. Some test names appear in lowercase in test.log but in uppercase in fix.log. We normalize all test names to lowercase to ensure alignment.
    \item Unstable test identifiers. Some test names are dynamically generated with timestamps or random suffixes, making them non-deterministic. Such instances are excluded.
    \item Log interleaving in Java. In some Java projects, test outputs from concurrent threads are interleaved without delimiters, making rule-based log parsing infeasible. This is likely due to unsynchronized multi-threaded logging.
\end{itemize}



\section{Conclusions and Future Works}
We introduce \multiswebench, a multilingual benchmark for issue resolving, consisting of $1,632$ human-validated GitHub instances on $7$ widely used programming languages. 
Based on this benchmark, we evaluate nine popular models using three agent methods and conduct a thorough analysis of the results.
Additionally, we establish the \multiswerl open-source community, aimed at building large-scale RL training datasets for issue-resolving tasks.
To catalyze community involvement, we release $4,723$ validated instances along with the complete data construction pipeline, encouraging the open-source community to continuously contribute and expand the dataset.
Looking ahead, we plan to scale \multiswebench and \multiswerl to more instances, languages, and modalities.
Beyond issue resolving, we would like to incorporate a broader range of software engineering tasks into our benchmark and RL community,
such as end-to-end project generation~\citep{zan2024codes,paperbench}, runtime environment setup~\citep{env_3,env_1,env_2}, bug reproduction~\citep{bug_2,bug_3} and localization~\citep{bug_1}, and software testing and maintenance~\citep{test_1,test_2}.
Overall, we envision our efforts as a step toward establishing a scalable and sustainable data-centric infrastructure that empowers future research.

\newpage

\clearpage
\newpage

\section*{Contributions}
\label{sec:contributions}

\setlength{\parindent}{0em}
\begin{multicols}{2}
\small
\creditsectionheader{Project Lead}
\corecontributor{Daoguang Zan$^\star$}
\\
\creditsectionheader{Core Contributors}
\textbf{Code Development}\\
\corecontributor{Zhirong Huang$^\star$}
\corecontributor{Hanwu Chen}
\corecontributor{Daoguang Zan}
\\
\textbf{Paper Writing}\\
\corecontributor{Wei Liu}
\corecontributor{Daoguang Zan}
\\
\textbf{Data Construction}\\
\corecontributor{Zhirong Huang}
\corecontributor{Hanwu Chen}
\corecontributor{Linhao Zhang}
\corecontributor{Daoguang Zan}
\\
\textbf{Agent Integration and Evaluation}\\
\corecontributor{Shulin Xin}
\corecontributor{Lu Chen}
\corecontributor{Qi Liu}
\corecontributor{Xiaojian Zhong}
\corecontributor{Aoyan Li}
\corecontributor{Daoguang Zan}
\\
\textbf{Result Analysis and Statistics}\\
\corecontributor{Wei Liu}
\corecontributor{Aoyan Li}
\corecontributor{Shulin Xin}
\corecontributor{Daoguang Zan}
\\
\creditsectionheader{Other Contributors}
\textbf{Leaderboard Refinement}\\
\corecontributor{Siyao Liu}
\corecontributor{Linhao Zhang}
\\
\textbf{Outsourced Hiring}\\
\corecontributor{Yongsheng Xiao}
\\
\textbf{Discussion and Support}\\
\corecontributor{Liangqiang Chen}
\corecontributor{Yuyu Zhang}
\corecontributor{Jing Su}
\corecontributor{Tianyu Liu}
\corecontributor{Rui Long}
\\
\creditsectionheader{Team Management}
\corecontributor{Kai Shen}
\corecontributor{Liang Xiang}
\\
\end{multicols}

Names marked with $^\star$ denote equal contribution.

\section*{Acknowledgments}
We gratefully acknowledge all members of the Seed-Foundation-Code team.
We thank Dong Chen, Ailun Yu, Shaoxin Lin, Yifan Shi, Bo Shen, Guangtai Liang, and Qianxiang Wang, former colleagues of Daoguang at Huawei Cloud, for their early discussion.
We thank Changxin Pu and Xiang Gao at Bytedance for their support with data annotation.
We thank Professors Jinxi Li from Shenzhen Technology University; Wei Zhang, Haiyan Zhao, and Zhi Jin from Peking University; and Xudong Lu and Lizhen Cui from Shandong University.
We thank Shulin, Wei, Aoyan, Lu, and Qi for their dedication in the final sprint before the deadline.
We are especially grateful to Zhirong, Hanwu, Linhao, and Daoguang for their countless late nights devoted to developing the dataset, without which this work would not have been possible.

\newpage

\bibliography{main.bib}

\newpage

\appendix

\end{document}